\documentclass{aa}
\usepackage{hyperref}
\hypersetup{
    unicode=false,          
    colorlinks=true,       
    linkcolor=blue,          
    citecolor=blue,        
    filecolor=blue,      
    urlcolor=blue           
}

\usepackage{threeparttable}
\usepackage{graphicx}
\usepackage{adjustbox}
\usepackage{txfonts}
\usepackage{lipsum}
\usepackage{booktabs}
\usepackage{xcolor}

\usepackage{subcaption}         
\usepackage{lscape}             
\usepackage{placeins}           
\usepackage[version=4]{mhchem}     

    \newcounter{maacounter}
    \newenvironment{maaenvironment}[1][]{\refstepcounter{maacounter} \nobreakspace 
   {\color{orange}{MAA(\themaacounter):~{#1}}} \rmfamily}{}
    
    \newcommand{\insr}[1]{\begingroup\color{olive}#1\endgroup}
    \usepackage[normalem]{ulem} 

\newcommand{\Aenus}{\texttt{Aenus-ALCAR}\xspace}
\newcommand{\GENEC}{\texttt{GENEC}\xspace}
\newcommand{\MESA}{\texttt{MESA}\xspace}
\newcommand{\ag}{\texttt{G20}\xspace}
\newcommand{\ad}{\texttt{M13}\xspace}

\defcitealias{Griffiths2026PaperI}{Paper I}
\defcitealias{Griffiths2026PaperII}{Paper II}
\let\linenumbers\relax
\let\nolinenumbers\relax
\begin{document}

   \title{The first 3D MHD core-collapse progenitors I:}

   \subtitle{General properties, convection and nuclear burning}

%

   \author{A. Griffiths\inst{1,2}
        \and Miguel-{\'A}. Aloy\inst{1,3}
        \and M. Obergaulinger\inst{1,3}
        }

 \institute{Departament d'Astronomia i Astrofísica, Universitat de València, 46100 Burjassot, Spain
 \and
Astrophysics Group, Lennard-Jones Laboratories, Keele University, Keele ST5 5BG, UK
\\
 \email{a.griffiths@keele.ac.uk, miguel.a.aloy@uv.es, martin.obergaulinger@uv.es}
\and
Observatori Astronòmic, Universitat de València, 46980 Paterna, Spain}
\date{Received September 30, 20XX}

 
  \abstract
   {The most energetic core-collapse supernovae are thought to arise from rapidly rotating, magnetised progenitors, yet the three-dimensional structure of their pre-collapse interior remains poorly constrained, and realistic distributions of magnetic fields, angular momentum, and convective asphericities are still lacking.}
   {We construct physically consistent three-dimensional pre-supernova progenitors including rotation and magnetic fields. In this first paper, we focus on the behaviour of turbulence and nuclear burning in the shells surrounding the stellar core, and assess their deviations from one-dimensional stellar-evolution models. The magnetorotational properties of the progenitors are discussed in the second paper of this series.} 
   {We used \Aenus to perform three-dimensional magnetohydrodynamic (MHD) simulations of two compact Wolf--Rayet progenitors obtained from the stellar evolution codes \GENEC and \MESA. The models were mapped into the multidimensional domain several minutes before collapse and evolved until the onset of core collapse.
   }
   {We find that in extended oxygen-burning shells, turbulent velocities exceed the standard mixing-length-theory (MLT) predictions by approximately a factor of two. In contrast, a thin silicon-burning shell is poorly described by MLT: mixing is reduced near both shell boundaries, and the inferred effective diffusion profile departs significantly from the standard one-dimensional prescription. These differences directly affect the spatial extent and efficiency of nuclear burning. In one of our models, a late shell merger present in the 1D stellar-evolution calculation does not occur in the 3D MHD evolution, indicating that such events may be sensitive to the adopted modelling framework.} 
   {We present the first 3D MHD pre-supernova progenitors of this kind, suitable for subsequent collapse and explosion calculations, and show that multidimensional effects can significantly modify turbulent mixing and shell burning during the final stages of massive-star evolution. We propose prescriptions to account for these effects in the advanced phases of stellar evolution.}


   \maketitle

\section{Introduction}


Massive stars with initial masses approximately between 8 $M_{\odot}$ and 60 $M_{\odot}$ end their lives as core-collapse supernova (CCSNe; see \citealt{Hirschi_2025} for a recent review). While many such explosions are thought to be powered by  neutrino heating \citep{Colgate_1966}, the most energetic events---including hypernovae and some long gamma-ray bursts---likely require rapidly rotating, strongly magnetised progenitors \citep[e.g.][]{Woosley_Bloom_2006,Nomoto_Tanaka_Tominaga_Maeda_Mazzali_2007,Burrows_2007,Mueller_2024arXiv240318952}. The current status of the theoretical and computational advances of such explosions is reviewed in \cite{Janka_2012,Muller_2020,Burrows_Vartanyan_2021} showing that
realistic predictions for such explosions depend critically on the pre-collapse configuration of angular momentum, magnetic fields, and convective asphericities in the stellar interior.

Magnetorotational explosions may also provide favourable conditions for the production of heavy $r$-process nuclei \citep[e.g.][]{Reichert_2022,Zha_muller_Powell_2024}, and both the explosion dynamics and the nucleosynthesis have been shown to depend sensitively on the magnetic-field configuration of the progenitor \citep{Obergaulinger_2017MNRAS.469L..43, Obergaulinger2020, Obergaulinger_2021, Aloy_2021, Bugli_2020,Bugli_Guilet_Obergaulinger_2021,Reichert_2024}.

The initial conditions for CCSNe simulations are crucial for understanding which progenitors explode and the nature and geometry of the resulting explosion. Over the past decade, increasing effort has focused on constructing multidimensional pre-supernova (pre-SN) progenitor models evolved during the final minutes preceding core collapse \citep[e.g.,][]{Couch_Chatzopoulos_Arnett_Timmes_2015,2016_muller,Yoshida_2021}. These models have proved valuable for assessing the impact of pre-collapse asymmetries on the neutrino-driven mechanism and have already been used as initial conditions for CCSNe simulations \citep{Muller_Melson_Heger_Janka_2017,Bollig_2021, Varanyan_2022MNRAS.510.4689}. However, most of these models neglect rotation or magnetic fields and were primarily designed to address the impact of asphericities on the neutrino-driven mechanism  (see \cite{Muller_Janka_2015} for a detailed analysis of the impact of symmetry breaking in CCSNe). They therefore do not provide realistic initial conditions for magnetorotational explosions as even the available rotating multidimensional progenitors remain non-magnetised \citep{Fields_2022}.

Whilst the importance of progenitor magnetic-field topology and strength for magnetorotational explosions is clear, fully self-consistent 3D rotating magnetic progenitors remain lacking, and most magnetorotational explosion studies are initialised from 1D progenitors \citep[e.g.][]{Mosta_Richers_Ott_Haas_Piro_Boydstun_Abdikamalov_Reisswig_Schnetter_2014,Obergaulinger2020}. 
This limitation arises due to current one-dimensional (1D) stellar-evolution (SE) calculations remaining intrinsically local in nature. Indeed, SE calculations consider magnetic fields through their impact on angular-momentum and chemical transport and are often treated using prescriptions based on the Tayler–Spruit (TS) dynamo \citep{Spruit_2002}, with more recent variants proposed by, e.g., \citet{Fuller_Piro_Jermyn_2019} and \citet{Eggenberger_2022}. In some cases, the possible role of the magnetorotational instability (MRI; \citealt{Balbus_Hawley_1991}) is also included \citep{Wheeler_Kagan_Chatzopoulos_2015,Griffiths_2022}. The predicted fields in a SE model are then based on the saturation field of the instability. This approach can yield approximate field strengths in radiative regions, but neither the global magnetic geometry of the star nor the magnetic connectivity across radiative and convective shells is captured.

Existing multidimensional MHD studies have shown that magnetic fields can persist within convective burning shells prior to collapse \citep{Varma_2021,Varma_2023}, but they do not yet provide fully self-consistent whole-star pre-collapse MHD progenitors suitable for collapse calculations. In particular, calculations that excise the core cannot determine how the field connects the iron core to the convective shells and outer layers. 


Uncertainties in pre-SN progenitors are not limited to magnetic fields and rotation. One-dimensional SE models also describe turbulent shell convection through effective prescriptions, chiefly mixing-length theory (MLT; \citealt{Bohm_1958,Kippenhan_1990}). Although such a treatment is unavoidable in long-term stellar evolution \citep[see][for a modern review on the use of MLT in 1D SE modelling]{Joyce_2023}, its validity in the advanced burning stages remains uncertain, since the structure, turnover times, and burning conditions of the final convective shells differ substantially from those of earlier evolutionary phases. Multidimensional simulations can therefore play a dual role: they can provide initial conditions at collapse and help test the one-dimensional treatment of turbulent mixing and nuclear burning in the final phases. In this respect, previous multidimensional studies have already proved useful for assessing and calibrating aspects of MLT in deep stellar interiors \citep[e.g.][]{Meakin_Arnett_2007,Jones_2017, Arnett_2019, Georgy_2024}. 

This first paper (\citetalias{Griffiths2026PaperI}, hereafter) focuses on the behaviour of turbulence and nuclear burning in the shells surrounding the stellar cores of our 3D progenitors, with the aim of assessing how their multidimensional behaviour differs from the corresponding 1D SE description during these late evolutionary phases. This analysis is intended to provide guidance for improving 1D SE modelling. The magnetic-field and angular-momentum properties of the models will be discussed in the second paper of this series (\citetalias{Griffiths2026PaperII}).

The paper is structured as follows. Section~\ref{sec:prog_model} presents the progenitor models and the key aspects of their prior SE. Section~\ref{sec:methods} describes the numerical methods, the initialisation of our 3D MHD models, and the nuclear reaction network used in our calculations.  Section~\ref{sec:results} presents the results, with particular emphasis on turbulence in the convective regions and the impact of 3D MHD modelling on nuclear burning compared with 1D SE calculations. We then discuss the implications of our models in Section~\ref{sec:discussion} and how the results may be applied to improve CCSNe progenitor modelling. Finally,  Section~\ref{sec:conlusions} summarises our main findings.

\section{Progenitor models}

\label{sec:prog_model}




In this work, we consider two progenitors. The first is the  $13\,M_{\odot}$ model of series B of \citet{Aguilera-Dena_2020}, computed with \MESA \citep[v.10398][]{Jermyn_2023} and referred to here as \ad. The second is a $20\,M_{\odot}$ model computed specifically for this work using the latest version of \GENEC \citep{Griffiths_2025}, referred to as \ag. Although each model follows different evolutionary histories and were produced with different SE codes, they both reach compact pre-collapse configurations of stripped-envelope, Wolf--Rayet-like stars. At the same time, they differ in their rotational history and in the detailed structure of their burning shells, making them useful contrasting cases for the present 3D MHD study. These progenitors may produce highly energetic and asymmetric magnetorotational supernova explosions and, likely, power superluminous supernovae or long gamma-ray bursts \citep[GRBs; indeed, the post-bounce and explosion of the 2018 version of model \ad, from \citet{Aguilera-Dena_2018}, has already been computed in axial symmetry by][]{2022_Obergaulinger}.





The two models differ not only in their initial mass, but also in their metallicity and initial rotation rate.
Model \ag has an initial metallicity of $0.1\,Z_{\odot}$ and a ZAMS surface rotation velocity of $460 \rm \, km \, s^{-1}$, whereas \ad has an initial metallicity of $0.02\,Z_{\odot}$ and a ZAMS surface rotation velocity of $600 \rm \, km \, s^{-1}$.
In model \ad, magnetic effects are treated through the TS-dynamo using the prescription of \cite{Heger_Woosley_Spruit_2005}. In model \ag we adopt the TS prescription
of \cite{Eggenberger_2022}.\footnote{The prescription of \citeauthor{Eggenberger_2022} employs a calibration parameter, and a power-law index for which we adopt $C_T=216$ and $n=1$, respectively. This parametrisation corresponds to the same physical dynamo description originally introduced by \citet{Spruit_2002}.} 
In addition, \ad includes an enhancement of rotational mixing by a factor 10 in order to induce chemically homogenous evolution (CHE) during the main sequence and helium burning phases, whereas \ag does not. Further details of the SE setup can be found in \cite{Aguilera-Dena_2020} and \cite{Griffiths_2025}, and, for the mass-loss treatment adopted in \ag, in \cite{2012_Ekstrom}.

During their lifetimes, both models have lost their hydrogen shells, and \ad has also lost its helium shell through earlier episodes of (strong) mechanical mass loss. They therefore end their lives as compact, hot, luminous Wolf--Rayet stars, with log$( L_{\rm} \ [L_{\odot}])> 5.25$, and surface temperatures, log$( T_{\rm eff} \ [K])> 5.2$. Their compactness is advantageous for the present study, since it allows the multidimensional simulations to encompass almost the entire star within a reasonably resolved computational domain.


The models are evolved until they approach core collapse. In the case of \ad, collapse is defined by the pre-SN link, that is
when $v_{\rm infall} = 10^8\, \rm cm \ s^{-1}$. For \ag, however, the lack of an acceleration term in \GENEC prevents an equally unambiguous identification of the pre-SN link. The model referred to here as ``collapse'' is the model that has the same central temperature as model \ad at the pre-SN link \citep[see][for further discussion]{Griffiths_2025}.
To initialise the multidimensional calculations, we select snapshots lying several minutes before collapse on the basis of exploratory 1D hydrodynamic runs.
The mapping times are indicated by the red dashed lines in Fig.~\ref{fig:Kipp_13M_20M}.

\begin{figure}[t!]
    \centering
    \includegraphics[width=\columnwidth]{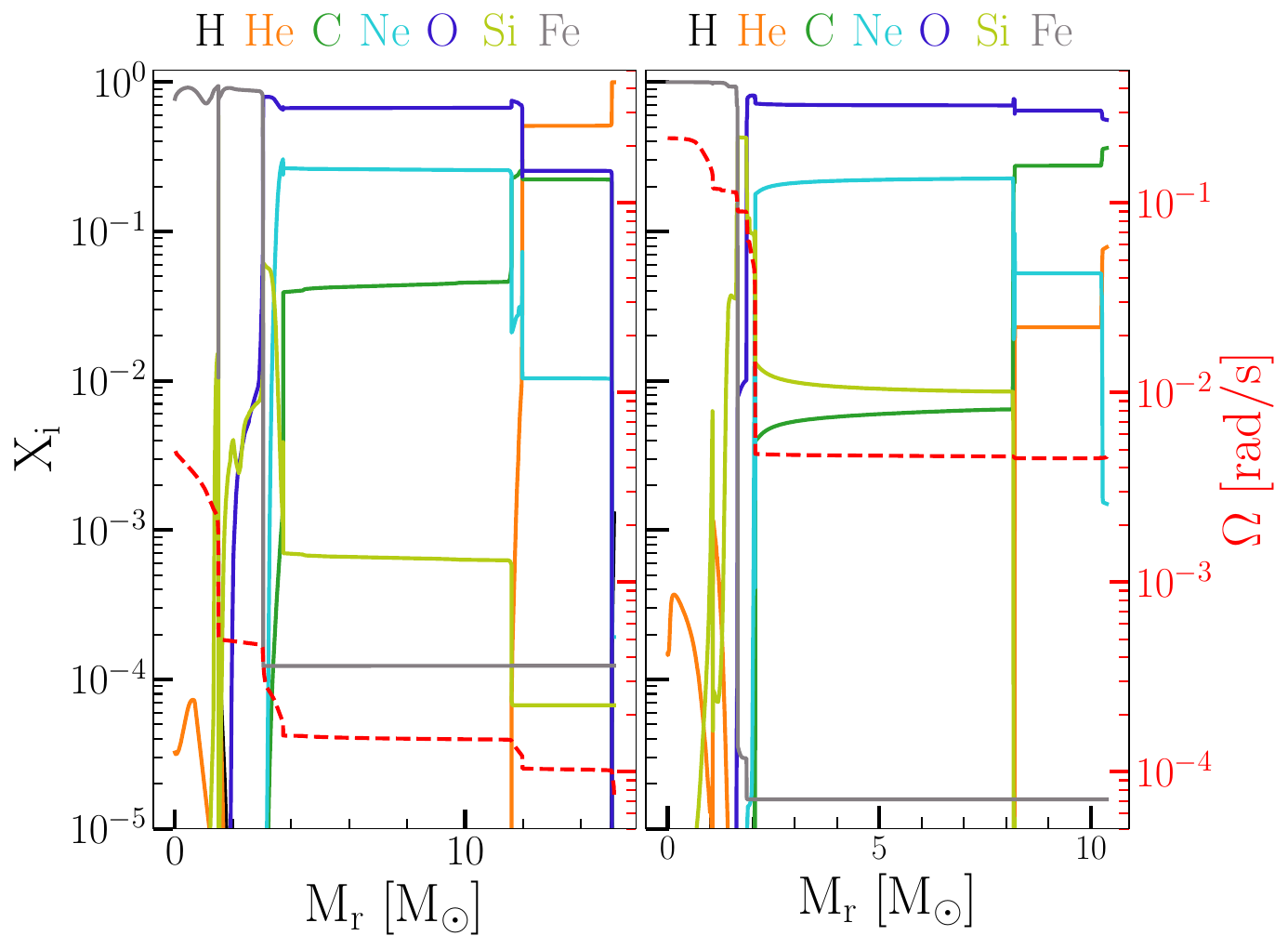}
    \caption{Chemical structure for key species at mapping for \ag (left), and \ad (right). The rotational frequency profile, $\Omega$ (red dashed line), is scaled according to the right axis.}
    \label{fig:Struc_at_fin}
\end{figure}

The chemical structure and rotation profiles at mapping are shown in Fig.~\ref{fig:Struc_at_fin}. Both progenitors exhibit extended oxygen-rich shells above the iron core, but their inner shell structure differs markedly. Model \ad retains a thin silicon-burning shell, whereas in \ag the silicon shell has almost entirely disappeared: most of it has already been processed into iron-group material, and the remainder has been mixed into the overlying oxygen shell. 
As a result, the iron core is more extended in \ag, reaching beyond $2.5\,M_{\odot}$. 
%
%
These differences foreshadow the contrasting multidimensional behaviour discussed later, in particular the presence of a thin Si-burning shell in \ad and a more extended O-burning structure in \ag.

Both models are divided into a radiative iron-core, a convective turbulent shell followed by a radiative shell and finally a large convective oxygen-burning shell above.
In the case of \ad the first convective shell is a thin Si-burning shell as mentioned, whereas for \ag it is the edge of the iron-core, in the process of photodisintegration, which is turbulent. At mapping the central core conditions of both models are similar, although model \ag is more compact than \ad and has a smaller value of $M_4$ (Eq.~\eqref{eqn:M4}).

A summary of the pre-collapse evolution can be found in the appendix~\ref{sec:precollapse} and a broader set of quantities characterising the two progenitors at mapping and at collapse can be found in appendix~\ref{sec:appexdix_table}.

\section{Methods and initialisation}

\label{sec:methods}

The 1D SE snapshots described in Sec.~\ref{sec:prog_model} are mapped into the multidimensional neutrino-MHD code \Aenus \cite{2015_Just}. Although the code has been widely used in core-collapse supernova simulations \citep{Obergaulinger2020,Aloy_2021,Obergaulinger_2021}, the present pre-collapse problem requires a dedicated numerical setup and initialisation strategy. In this section, we describe the numerical methods adopted here, the procedure used to maintain hydrostatic equilibrium after mapping, the initialisation of the convective flow, and the reduced nuclear reaction network employed in the multidimensional calculations.

\subsection{Numerical methods}

The MHD equations are solved  in spherical coordinates, $(r, \theta,\phi)$, using a finite-volume discretization. The numerical methods employed follow \citet{Obergaulinger2020}: we use high-resolution shock-capturing techniques combining high-order reconstruction and approximate Riemann solvers (specifically HLLC). Unlike in post-bounce CCSN simulations, no neutrino transport is included here, since densities in the pre-collapse evolution remain far below $10^{11} \, \rm g \ cm^{-3}$. We do account for thermal neutrino losses through the effective prescription of \citet{Itoh_Hayashi_Nishikawa_Kohyama_1996}, as in SE calculations, and we also include neutrino losses from weak reactions within the nuclear reaction network.

%
The production 3D simulations use a spherical polar grid with
$(n_r,n_{\theta},n_{\phi})=(640,128,256)$. The radial grid is logarithmically spaced, with the centre of the innermost cell located at $7.5\,\mathrm{km}$. The outer boundary is placed at $2\times10^5\,\mathrm{km}$ for \ag and at $10^5\,\mathrm{km}$ for \ad, in both cases beyond the main oxygen shell.\footnote{The outer radial boundaries are chosen so that they lie beyond the main oxygen shell in each model.}

The exact equations solved are given in Appendix~\ref{sec:MHD_eqs}. We explicitly neglect thermal diffusivity in the numerical calculations. This approximation is justified by the large value of the Péclet number in all convective regions. The latter is defined as
\begin{equation}
\label{eqn:Peclet}
    \mathrm{Pe} = \frac{3D_{\rm MLT}}{\kappa}.
\end{equation}
Here $\kappa$ is the thermal diffusivity and $D_{\rm MLT}$ is the mixing-length diffusion coefficient, $D_{\rm MLT} = v_{\rm conv} H_p/3$,  where $H_p$ is the pressure scale height,%
\footnote{Only an order-of-magnitude estimate is required here, so we simply take the pressure-scale height, $H_p$ as the characteristic mixing length scale in MLT.} and $v_{\rm conv}$ is the standard MLT estimate of the convective velocity

\begin{equation}
\label{eqn:v_conv}
    v^2_{\rm conv} = g \beta (\Delta\nabla) \frac{\Lambda^2_{\rm MLT}}{8H_p},
\end{equation}
where $\beta=\left(\frac{d\ln\rho}{d\ln T}\right)_P$, $\Delta\nabla$ is the super-adiabatic gradient, and $g$ the gravitational acceleration. Equation \eqref{eqn:v_conv} expresses the balance between the work done by buoyancy and the kinetic energy acquired by the eddy. 

For all convective regions considered in this work $\mathrm{Pe}\sim10^7$ (see Tab.~\ref{tab:convective_zones}), implying that thermal diffusion acts on timescales much longer than the convective turnover times in our models.

\subsection{Maintaining hydrostatic equilibrium}

Mapping 1D SE models into a multidimensional Eulerian code inevitably introduces small departures from hydrostatic equilibrium. This can occur due to the SE snapshots providing cell-centred quantities, whereas the MHD code evolves cell-averaged variables, and because the discretisation and numerical solvers used by the two codes differ substantially. 
To minimise the resulting transient motions, we apply the hydrostatic correction method outlined by \cite{Zingale2002}. 

Our implementation adjusts both the density ($\rho$) and pressure ($P$) profile%
\footnote{In \citet{Couch_Chatzopoulos_Arnett_Timmes_2015} the authors also employ the method of \cite{Zingale2002} when initialising their 3D simulations, but they keep  the pressure profile fixed while adjusting only the density.} by means of a bisection procedure so as to improve the numerical accuracy of the hydrostatic-equilibrium equation,
\begin{equation}
\label{eqn:hydro_zing}
    \frac{dP}{dr} = \rho g.
\end{equation}
We assume the temperature, $T$, and chemical composition, $X_i$, to remain fixed and close the system with the equation of state (EoS). In the original method, $g$ is kept fixed during the correction. Here, however, the modified density profile alters the gravitational field. We therefore proceed iteratively: after each correction step, we recompute the gravitational potential, $\Phi$, from the Poisson equation,
\begin{equation}
\label{eqn:Poisson}
    \nabla^2 \Phi = 4\pi G \rho(r),
\end{equation}
and repeat the correction until the relative change in the value of g between successive iterations falls below an arbitrary tolerance $\varepsilon=10^{-12}$. On average, $\sim 10$ iterations are sufficient to achieve this level of convergence. With relative changes of at most 3\% in the density and pressure profiles, we improve the accuracy of Eq.~\eqref{eqn:hydro_zing} by three orders of magnitude.

\begin{figure}[t!]
    \centering
    \includegraphics[width=\columnwidth]{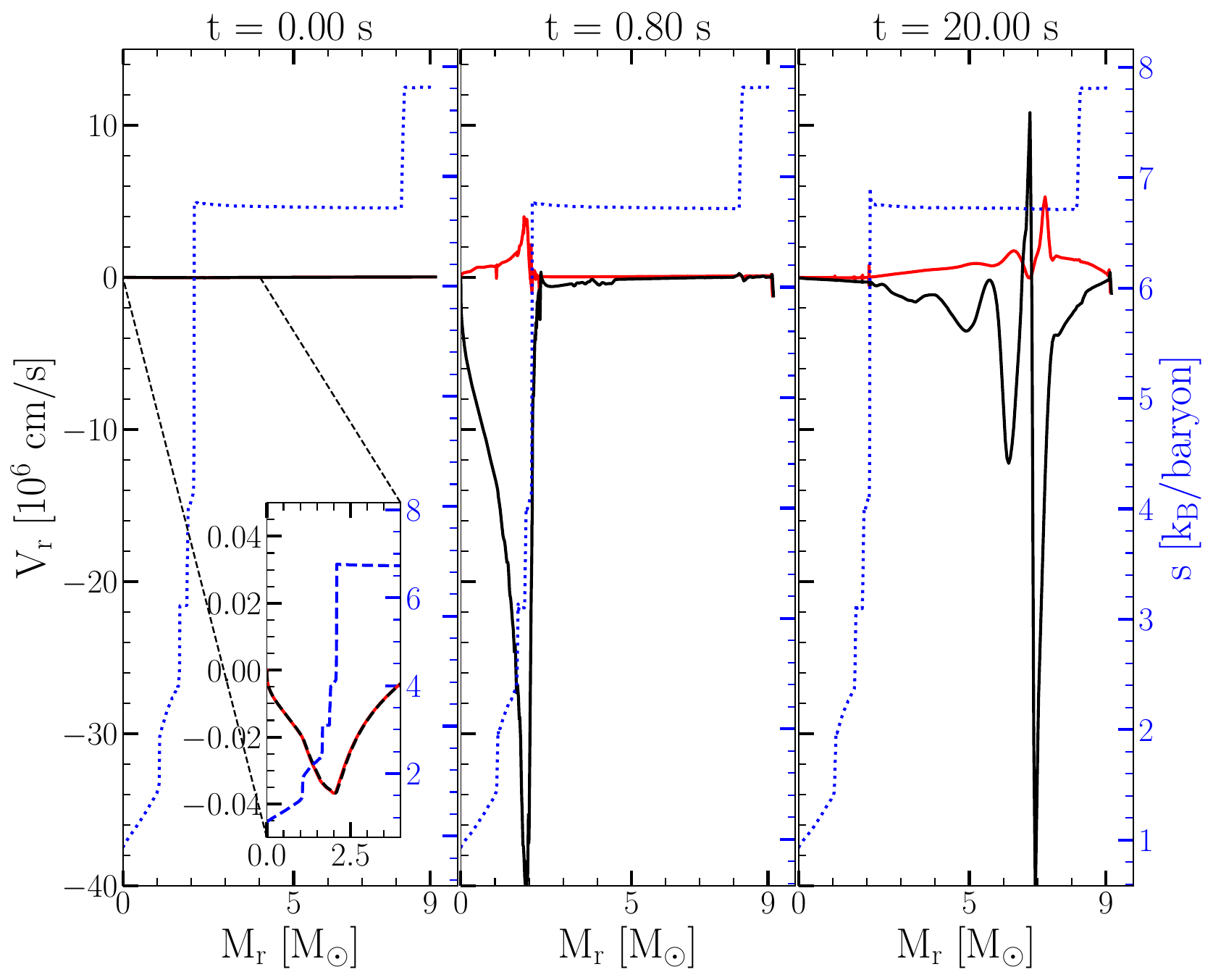}
    \caption{Radial velocity profiles of the corrected (red) and non-corrected (black) stellar data of model \ad at $0$ seconds (left), 0.8 seconds (centre) and 20 seconds (right) after mapping. The entropy profile of the corrected model is also shown (blue dotted line). The inset displays a zoom of the infall velocity inherited from the stellar data. }%
    \label{fig:Vx_corr}%
\end{figure}

To illustrate the effect of this correction, we performed 1D hydrodynamic simulations in \Aenus using both the original, untouched, SE profiles and the corrected ones. Figure~\ref{fig:Vx_corr} shows the radial velocity as a function of mass for model \ad at three different times after mapping. The corrected model exhibits much smaller transient sound waves and, after $\sim 20\,$s the radial velocity is nearly zero--and, crucially far below the SE-predicted turbulent velocities--throughout almost the entire star. Applying the correction method therefore ensures that no artificial radial waves interfere with the convective or rotational dynamics.

\subsection{Initialising the convective flow}

Multidimensional models require a transition phase before the convective regions develop fully turbulent flow. Depending on the size of the region, this phase may last from a few seconds to several tens of seconds. As a first estimate, we characterise the local convective turnover time from the SE model as
\begin{equation}
    \tau_{\rm conv, SE} = \frac{2 H_p}{v_{\rm conv}}.
    \label{eq:tau_conv}
\end{equation}
%
To significantly reduce the duration of the transient phase, we impose an initial velocity perturbation in the $r$ and $\theta$ components of the flow, restricted to the convective regions. 



To construct this initial perturbation, we first performed exploratory 2D models without any imposed perturbation and allowed the flow to develop naturally. After many turnover times, once convection has fully developed, the angular power spectrum of the velocity field approaches a statistically stationary form. We then use the resulting decomposition as a guide to build a synthetic perturbation. In practice, we find that the leading term in the spherical-harmonic decomposition peaks around $\ell_{\rm peak} = 5$, while the low-$\ell$ part of the spectrum rises approximately as $\ell\propto 3/2$, reminiscent of a Kazantsev-like form, before transiting to a Kolmogorov slope, $\ell \propto -5/3$. 
%
We therefore construct a synthetic initial power spectrum with this shape and  normalise it locally to the MLT velocity $v_{\rm conv}$ inferred from the SE data, so that the perturbation amplitude reflects the local convective strength. 

To recover the full velocity field, we assign phases to the spectral components using the phase pattern extracted from the converged 2D models.%
\footnote{The essential requirement for spectral convergence is that the initial phase distribution does not contain any privileged directions, for instance aligned with the poles.
} 
The final turbulent spectra obtained in the perturbed and unperturbed models are consistent with one another, but the perturbed models reach their saturated turbulent state much faster. This procedure therefore reduces the fraction of simulation time spent in the initial transient regime and increases the effective duration of the fully developed turbulent phase.

\subsection{Nuclear reaction network}

A key ingredient of both the SE progenitors and the MHD simulations is the nuclear reaction network employed. The two fundamental requirements are to track the reduction of $Y_e$ in the core and the energy production in convective layers. Large networks are computationally prohibitive in multi-D simulations due to their memory and computation costs. For this reason, we adopt a reduced network in our MHD simulations, analogous to one used in the SE models.

The network used here is \texttt{RN28}, implemented through \texttt{RENET} \citep{Navo_2023}, a reduced version of the nuclear-reaction code \texttt{WinNet} \citep{WinNet_2023} available in \Aenus. In \cite{Navo_2023}, two networks were implemented in the code: \texttt{RN16}, a basic $\alpha$-chain up to $\ce{^56Ni}$, and \texttt{RN94}, which extends the network up to $\ce{^92Mo}$, including both proton- and neutron-rich nuclei. Neither network is optimal for the present problem. \texttt{RN16} is too limited to reproduce the reduction of $Y_e$ in the core as it becomes neutron-rich, whereas \texttt{RN94}, although more complete, is too expensive for the 3D simulations carried out here.

We therefore constructed \texttt{RN28}, a new reduced network that remains small enough for 3D simulations yet reproduces the $Y_e$ evolution predicted by \texttt{RN94} under pre-collapse conditions. The isotopes included in \texttt{RN28} are shown in Fig.~\ref{fig:networks}, together with those of \texttt{RN94}. 
%
%
%
 Above $T=5.5\,$GK, the network switches to a nuclear statistical equilibrium (NSE) treatment; below that threshold, the full reduced  network is evolved explicitly. Weak reactions are included, and the associated energy losses are removed from the system. 

As the SE networks employ approximate effective prescriptions for weak reactions,\footnote{In both \texttt{GeValNet25} and \texttt{approx21}, used in the SE models, the reduction of $Y_e$ is tracked through an effective electron-capture chain ending at $\ce{^56Cr}$.} the $Y_e$ evolution in the multidimensional calculations is not identical to that obtained in SE models. However, it is based on a more explicit treatment of weak interactions within the reduced network.  In appendix~\ref{sec:appendix_reac} we demonstrate the performance of \texttt{RN28} compared to \texttt{RN94} and the networks used in the SE calculations.

\section{Results}
\label{sec:results}

The progenitors are first evolved in 2D to allow sufficient time for sound waves to relax and the nuclear reaction network to stabilise. For \ag, this transition phase only lasts $2\,$s, after which the model is mapped to 3D by imposing uniformity in the $\phi$-direction. In the case of \ad, the transition time is longer--$\sim 200$\,s--as the snapshots available from \cite{Aguilera-Dena_2018} were not output frequently enough to start the multidimensional evolution precisely five minutes before collapse.\footnote{The optimal starting point requires roughly $500 \,$s to reach collapse in 2D, but this duration was too long to evolve in 3D given the computing resources available.} 
The preceding 2D relaxation runs use the same radial and polar resolution as our 3D models, assuming axisymmetry.

\begin{figure}[t!]
\centering\includegraphics[width=\columnwidth]{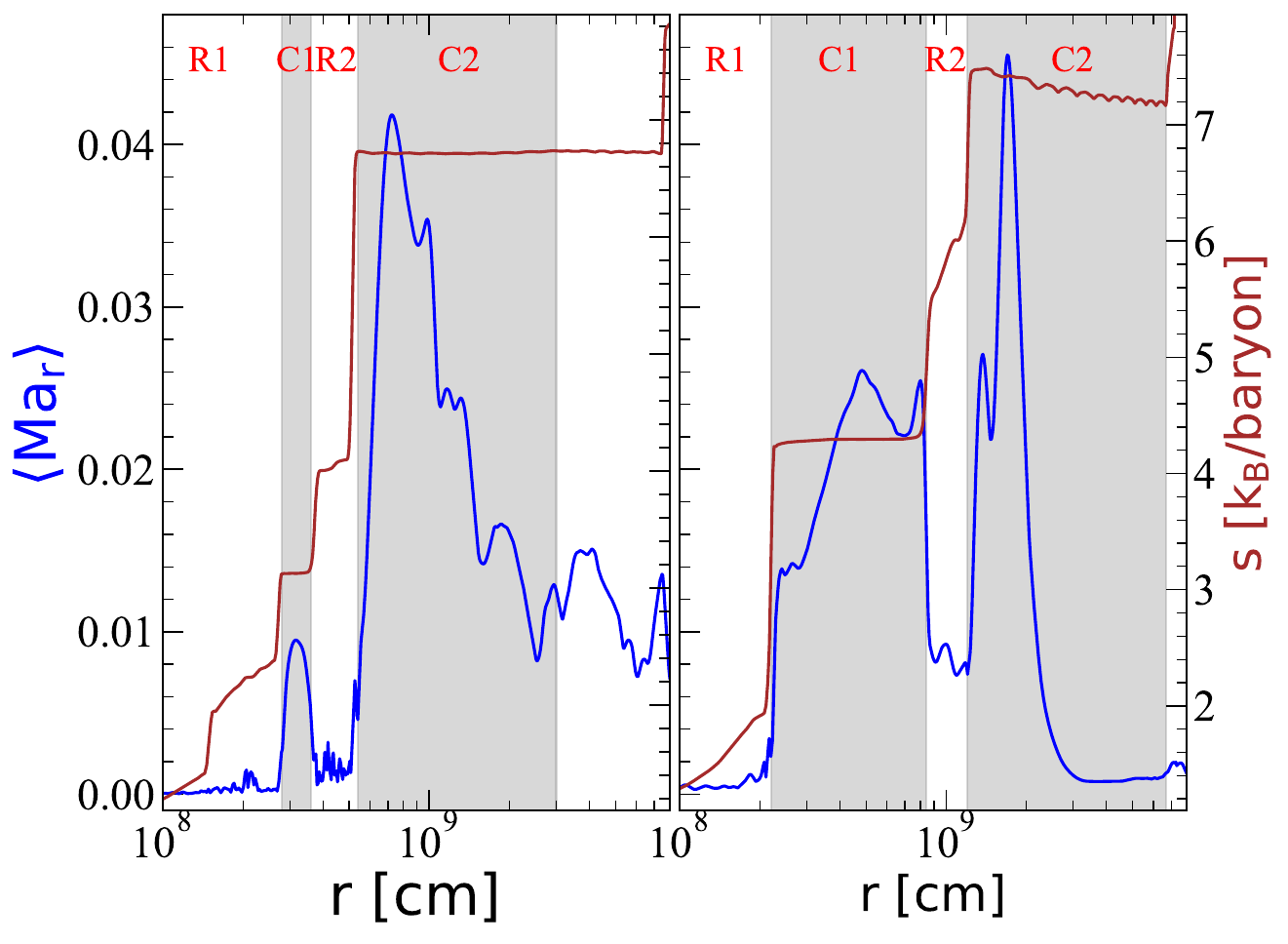} 
\caption{Angularly averaged turbulent radial Mach number, Eq.~\eqref{eq:Mach_r_turb3}, and entropy profiles after $60\,$s of 3D evolution for models \ad (left) and \ag (right). Grey shading highlights the convective regions. The labels R1, C1, R2 and C2 will be used to reference these zones in the text.}
\label{fig:Mar_entropy}
\end{figure}

%
The simulations cover several minutes of physical time before the onset of core collapse: 190\,s for \ag and 520\,s for \ad.
Each progenitor can be divided into four regions of interest: a radiative core, an inner convective shell, a radiative shell, and an outer convective shell. These regions are labelled R1, C1, R2, and C2, respectively, in Fig.~\ref{fig:Mar_entropy}, where we show the entropy profiles and the turbulent radial Mach number after $60\,$s of 3D evolution.
We define the turbulent average Mach number as
\begin{equation}
\label{eq:Mach_r_turb3}
\langle \mathrm{Ma}_{r}\rangle(r)
=
\left[
\frac{1}{4\pi}
\int_\Omega
\left(
\frac{v_r(r,\theta,\phi)-\langle v_r\rangle_\Omega(r)}
     {c_s(r,\theta,\phi)}
\right)^2
\,d\Omega
\right]^{1/2},
\end{equation}
%
where $\langle\cdot\rangle_\Omega$ denote averages over the spherical shell of radius $r$ and $c_s$ is the local sound speed. In both models, the convective regions are clearly identifiable by peaks in $\mathrm{Ma}_{r}$, and coincide with shells in which the entropy gradient is nearly flat or decreasing. 

The convective regions in the 3D models generally coincide with those flagged by the Schwarzschild criterion in the SE calculations, with one notable exception: the region C1 of \ag. This layer was convective in the SE simulation shortly before the start of the multidimensional evolution (corresponding to the silicon shell shown in Fig.~\ref{fig:Kipp_13M_20M}), but is no longer formally flagged as convective at the time of mapping. Nonetheless, it rapidly develops turbulent motions in the 3D simulation. Since this turbulence is no longer sustained by nuclear burning, however, it gradually weakens by the end of the evolution.

\subsection{Analysis of turbulent regions}

The physical properties of each convective region are listed in Table~\ref{tab:convective_zones}. Quantities indicated with an overline refer to spatial averages taken over a representative snapshot in the 3D run, as they remain roughly constant throughout the evolution. Quantities reported with a tilde are further averaged in time, since they evolve during the 3D evolution. 

In the case of region C2 of \ad, the quoted upper radius is not that of the entire oxygen shell--whose radial extent is $\approx 7.5 \times 10^9$\,cm--but rather the  maximum height reached by  the convective flows during the 3D simulations. Once the turbulent flow approaches saturation, the convective motions in this shell do not span the entire oxygen shell; instead, they extend only up to $\sim 3.5 \times 10^9$\,cm. This behaviour is not observed in the other convective regions, which do reach the full radial extent of their respective burning layers. All of the shells differ substantially in both geometric extent and thermal conditions, with two particularly notable cases: the very thin silicon-burning shell in \ad, and the very extended oxygen-burning shell in \ag.

\def\arraystretch{1.1}
\begin{table*}[t!]
\caption{Properties of convective regions for 3D models.}
\label{tab:convective_zones}
\centering
\begin{threeparttable}
\begin{tabular}{lccccc}
\toprule
Models & \multicolumn{2}{c}{\ag} & & \multicolumn{2}{c}{\ad} \\
\cmidrule(lr){2-3}\cmidrule(lr){5-6}
 & C1 & C2 & & C1 & C2 \\
\midrule
$( r_{\rm bottom} ; r_{\rm top} )$ [$10^8$\,cm]
& (2.2 ; 8.4) & (13 ; 67) & & (2.8 ; 3.8) & (5.4 ; 30)\tnote{\textcolor{blue}{(a)}} \\

$( M_{\rm bottom} ; M_{\rm top} )$ [$M_{\odot}$]
& (1.47 ; 3.03) & (3.70 ; 11.6) & & (1.65 ; 1.86) & (2.07 ; 4.15) \\

$(T_{\rm bottom} ; T_{\rm top})$ [GK]
& (3.4 ; 1.9) & (1.5 ; 0.40) & & (3.1 ; 2.4) & (1.9 ; 0.72) \\

\midrule
$\overline{s}$ [kb/baryon] & 4.3 & 7.3 & & 3.2 & 6.7 \\
$\overline{H}_p$ [cm] & $1.6\times10^8$ & $9.5\times10^8$ & & $7.3\times10^7$ & $6.2\times10^8$ \\

\midrule
$\tilde{v}_{\rm turb}$ [$10^6$\,cm\,s$^{-1}$] & 14 & 31 & & 6.7 & 12 \\
$\tilde{\tau}_{\rm turnover}$ [s] & 15 & 138\tnote{\textcolor{blue}{(b)}} & & 16 & 72 \\
Pe\tnote{\textcolor{blue}{(c)}} & $7\times10^7$ & $3\times10^6$ & & $5\times10^7$ & $6\times10^6$ \\

\midrule
Three most abundant species
& \ce{^{54}Fe}, \ce{^{56}Ni}, \ce{^{50}Cr}
& \ce{^{16}O}, \ce{^{20}Ne}, \ce{^{24}Mg}
&
& \ce{^{28}Si}, \ce{^{32}S}, \ce{^{40}Ca}
& \ce{^{16}O}, \ce{^{24}Mg}, \ce{^{28}Si} \\

$\varepsilon_{\rm peak}$ [erg\,g$^{-1}$\,s$^{-1}$]
& $-2.0\times10^{14}$ & $1.5\times10^{13}$ & & $7.8\times10^{14}$ & $3.9\times10^{14}$ \\
\bottomrule
\end{tabular}

\begin{tablenotes}[flushleft]
\footnotesize
\item \textcolor{blue}{(a)} The entire oxygen shell extends out to $7.5\times10^{9}$\,cm, but the maximum radius that becomes turbulent is roughly $3\times10^{9}$\,cm. In the case of \ag, the whole oxygen shell is turbulent; thus we report the full shell size.
\item \textcolor{blue}{(b)} The shell becomes fully turbulent only toward the end of the simulation. At that point, the convective turnover time decreases to $\sim 20\,$s, with peak velocities of $4\times10^7$\,cm\,s$^{-1}$. Because this fully developed state is reached late, we report a value averaged over the entire evolution, including the ramp-up phase.
\item \textcolor{blue}{(c)} Peclet number (Eq.~\eqref{eqn:Peclet}).
\end{tablenotes}
\end{threeparttable}
\end{table*}

\begin{figure}[t!]
\centering\includegraphics[width=\columnwidth]{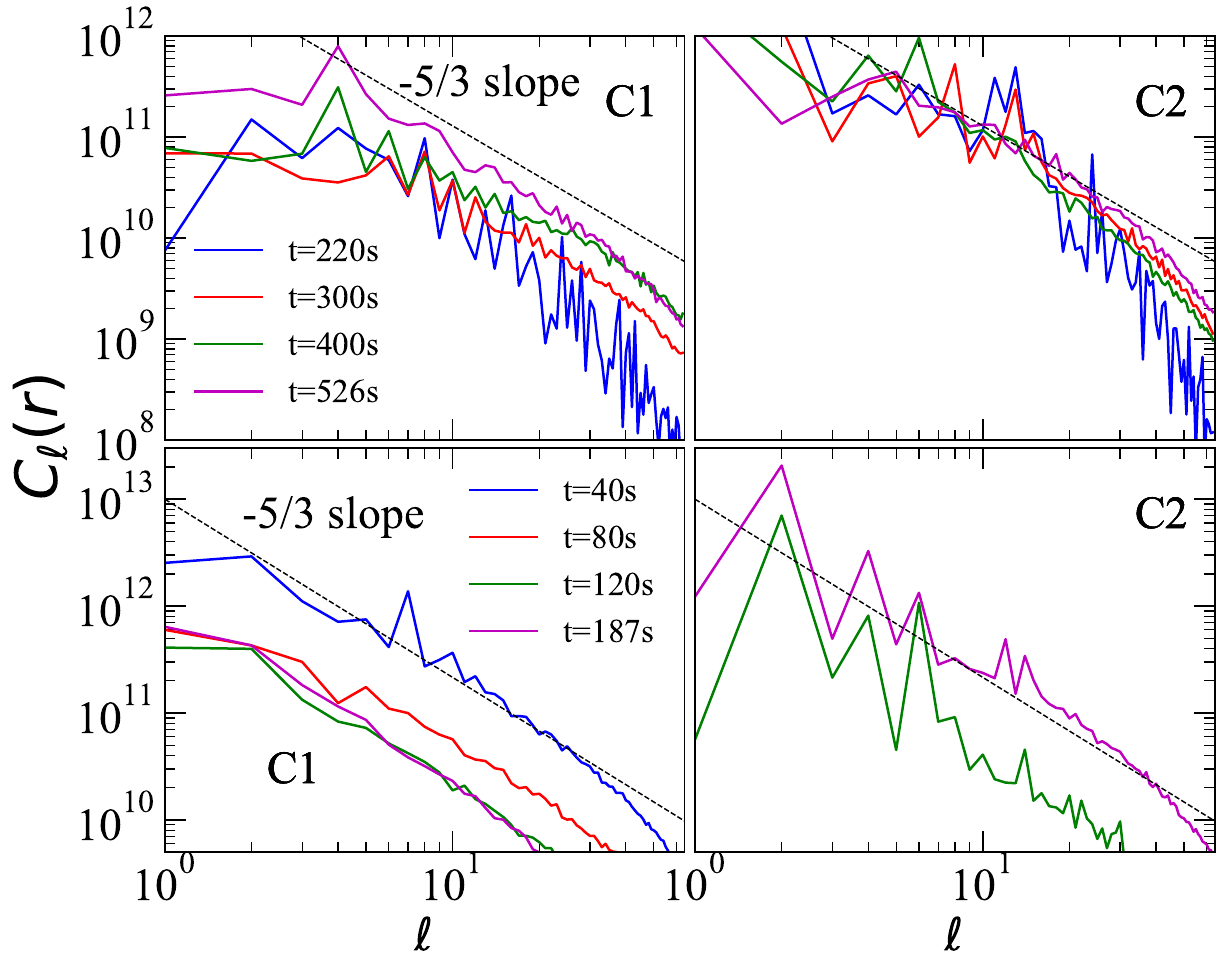} 
\caption{Power spectra of the spherical-harmonic decomposition of the turbulent velocity, Eq.~\eqref{eqn:turb_vel}, for \ad (top) and \ag (bottom). The inner convective regions (C1) are shown on the left, and the outer regions (C2) on the right. Dashed lines indicate a $\ell^{-5/3}$ scaling for reference.
}
\label{fig:spectra_13M_20M}
\end{figure}


%


We define the local turbulent velocity as
\begin{equation}
\label{eqn:turb_vel}
    v_{\rm turb}(r,\theta,\phi) =
    \left[
    \left(v_r-\langle v_r\rangle_{\Omega}\right)^2
    +v_\theta^2
    +\left(v_\phi-\langle v_\phi\rangle_{\Omega}\right)^2
    \right]^{1/2}.
\end{equation}
Typical turbulent velocities in the convective regions are of order $10^7\,\rm cm \, s^{-1}$, implying that fluid elements can travel approximately one scale height within tens of seconds. 

We estimate the local convective turnover time of each region as
%
\begin{equation}
\label{eq:tauturnoverloc}
    \tau_{\rm turnover,loc}(r) =
    \frac{\langle H_p\rangle_{\Omega}(r)}
         {\langle v_{\rm turb}\rangle_{\Omega}(r)},
\end{equation}
and the global turnover time of a convective region as the radial average between the bottom, $r_{\rm bottom}$, and top, $r_{\rm top}$, boundary of the shell
\begin{equation}
\tau_{\rm turnover} =
\frac{1}{r_{\rm top}-r_{\rm bottom}}
\int_{r_{\rm bottom}}^{r_{\rm top}}
\tau_{\rm turnover,loc}(r)\,dr.
\end{equation}

The resulting turnover times are $\sim15$\,s for the inner convective shells, allowing us to capture over 10 turnover times during the whole simulation. In the outer shells the turnover time is much longer (of order minutes), so we capture much fewer turnovers. Nonetheless, for \ad we still capture around five turnovers in C2, which is sufficient to estimate characteristic flow properties. For \ag, in contrast, the coverage of the outer convective  region is more limited (only a few turnovers are resolved during the 3D simulations). The average value reported in Table~\ref{tab:convective_zones} (138\,s) is influenced by the initial ramp-up phase, and therefore overestimates the turnover time of the shell once it is fully turbulent. Near the end of the simulation, the shell approaches $\tau_{\rm turnover}\simeq 20\,$s, indicating that the late-time flow is substantially more developed than the time-averaged value alone would suggest.

To characterize the geometry of the flow, we analyse the power spectra of  the spherical-harmonic decomposition of the turbulent velocity (Fig.~\ref{fig:spectra_13M_20M}), defined as
\begin{equation}
\label{eqn:C_ell}
    C_{\ell} ( r ) = \frac{1}{2\ell +1 } \sum_{m=-\ell}^{m=\ell} |a_{\ell,m}|^2 (r),
\end{equation} 
where,
\begin{equation}
\label{eq:a_lm}
    a_{\ell m} ( r )  = \int_{\Omega} f(r,\theta, \phi)\, Y_{\ell m}^*(\theta, \phi) \, d\Omega,
\end{equation}
and $Y_{\ell m}$ are the complex spherical harmonics.
Except for region C2 in \ag, the spectra have largely converged by the end of the simulations as shown by the close agreement between the last two snapshots in each panel of Fig.~\ref{fig:spectra_13M_20M}.

At large $\ell$, all spectra show a decay consistent with a Kolmogorov-like cascade. This decay sets in at lower $\ell$ in the outer convective shells (C2) than in the inner ones. Consequently, the characteristic angular scales, $\lambda \sim \pi R / \ell$, are larger in C2 than in C1. This trend is physically plausible, owing both to their larger average radii and to the lower values of $\ell$ at which the cascade begins. This shift may, however, be influenced by the reduced radial resolution in the outer regions resulting from the logarithmic grid. 

%
%

\subsection{Comparison to MLT}
\label{sec:comparisonMLT}

In 1D SE models, convection is described through MLT, in which buoyant eddies transport heat and chemical species over a characteristic length scale $\Lambda_{\rm MLT}=\alpha_{\Lambda_{\rm MLT}}H_p$. The corresponding effective convective diffusion coefficient is 
$$
D_{\rm conv,MLT}= \Lambda_{\rm MLT} v_{\rm MLT},
$$
with $\alpha_{\Lambda_{\rm MLT}}$ as the only free parameter. In both of the SE models considered here, $\alpha_{\Lambda_{\rm MLT}} = 1.5$.%
\footnote{Observational calibrations of this parameter can vary, for example with effective temperature; see \cite{Pinheiro_Fernandes_2013}.} 
Since this calibration is largely based on earlier evolutionary phases, multidimensional simulations of the final burning shells provide a useful test of its validity in the pre-collapse regime.

Following \cite{Meakin_Arnett_2007}, we compare the multidimensional temperature fluctuations  and turbulent velocities to the MLT expectations, $T'$ and $v_{\rm conv}$, respectively, in the convective regions. These quantities are related to the super-adiabatic gradient through two, a priori distinct, correlation coefficients:
\begin{equation}
\label{eq:Tprime_T}
\frac{T'}{T} = \alpha_T (\Delta\nabla),
\end{equation}
and,
\begin{equation}
\label{eq:v_conv_Meakin-Arnet}
v_{\rm conv} = \frac{\alpha_v}{2}\sqrt{g\beta (\Delta\nabla) H_p}.
\end{equation}

These relations imply two estimates for the effective mixing-length parameter, $\alpha_{\Lambda_{\rm MLT},T}$ and $\alpha_{\Lambda_{\rm MLT},v}$,  related to the above coefficients through $\alpha_v = \alpha_{\Lambda_{\rm MLT},v}/\sqrt{2}$ and $\alpha_T = \alpha_{\Lambda_{\rm MLT},T}/2$.

We compare the MLT prediction for the convective velocity (Eq.\,\eqref{eqn:v_conv}) with the radial turbulent velocity estimated in the 3D simulations by the root-mean-square (rms) velocity fluctuations,
\begin{equation}
\label{eqn:vturb_r}
    v_{{\rm rms},r}(r)
    =
    \left[
    \frac{1}{4\pi}
    \int_{\Omega}
    \left(v_r-\langle v_r\rangle_{\Omega}\right)^2
    d\Omega
    \right]^{1/2}.
\end{equation}
The ratio between $v_{\rm rms,r}$ and the SE value of $v_{\rm conv}$ provides the corresponding estimate of $\alpha_{\Lambda_{\rm MLT},v}$,
while the temperature fluctuations can be used to infer $\alpha_{\Lambda_{\rm MLT},T}$ through
\begin{equation}
\label{eq:ratio_vturb}
    \left(\frac{T'}{T}\right) \frac{1}{v^2_{{\rm conv}}} = \frac{ 4 \alpha_T}{\alpha^2_v g \beta H_p}.
\end{equation}

\def\arraystretch{1.1}%
\begin{table}[t!]
\centering
\caption[MLT parameters estimated from 3D simulations.]{MLT parameters estimated from simulations. All values correspond to spatial averages over each convective region once turbulence has reached saturation. The rms values refer to the root-mean-square deviations from the average value, whereas "up" and "down" denote the deviations between the mean upward or downward flows and the overall average.}
  \begingroup
  \setlength{\tabcolsep}{5pt} 
    \begin{tabular}{ c c c c c  }

        \hline 
        \hline
        Models & \multicolumn{2}{c}{\ag} & \multicolumn{2}{c}{\ad} \\\cline{2-2}\cline{4-5}
        & C2 & & C1 & C2 \\

       $\left\langle v_{{\rm rms},r} / v_{\rm SE}\right\rangle$ & 2.6  & & 0.89 & 1.8 \\
       $\left\langle v_{{\rm down},r} / v_{\rm SE}\right\rangle$ & 2.5 & & 0.60 & 1.4 \\
    $\left\langle v_{{\rm up},r} / v_{\rm SE}\right\rangle$ &  1.5 & & 0.60 &  1.5 \\
\noalign{\vskip 3pt}       
\hline
\noalign{\vskip 3pt}
       $\left\langle T_{\rm rms}' /T\right\rangle \times 10^4$ &  14 & & 10 & 3.9 \\
       $\left\langle T_{\rm down}' /T\right\rangle \times 10^4$ &  2.1 & & 4.9 & 2.5 \\
       $\left\langle T_{\rm up}' /T\right\rangle \times 10^4$ &  2.4 & & 7.5 &  0.96 \\
\noalign{\vskip 3pt}
\hline
\noalign{\vskip 3pt}
       $( 2 \alpha_T ; \sqrt{2}\alpha_v)_{\rm rms}$ & (3.8 ; 3.4)  & & (18 ; 1.4)&  (2.1 ; 2.8) \\
       $(2 \alpha_T ; \sqrt{2}\alpha_v)_{\rm up}$ & (0.59 ; 1.9) & & (9.0 ; 0.94) & (0.51 ; 2.3) \\
       $(2\alpha_T ; \sqrt{2}\alpha_v)_{\rm down}$ &  (0.64 ; 3.2) & & (14 ; 0.95) &  (1.4 ; 2.2) \\
\hline
\noalign{\vskip 3pt}
       $\left(\alpha_{\Lambda,T}/\alpha_{\Lambda,v}\right)_{\rm rms}$ & 1.1  & & 13 &  0.76 \\
       $\left(\alpha_{\Lambda,T}/\alpha_{\Lambda,v}\right)_{\rm up}$ & 0.32 & & 9.6 & 0.22 \\
       $\left(\alpha_{\Lambda,T}/\alpha_{\Lambda,v}\right)_{\rm down}$ &  0.21 & & 15 &  0.60 \\
\noalign{\vskip 3pt}       
\hline
\hline
\\
    \end{tabular}
    \endgroup
\label{tab:MLT_estimation}
\end{table}

To evaluate these coefficients, we follow the three approaches described in \citet{Meakin_Arnett_2007}. In the first, we compute the rms fluctuations of the 3D MHD models ($T'_{\rm rms}$ and $v_{\rm rms,r}$) to determine $\alpha_v$ and $\alpha_T$. In the second and third, we compute the same quantities separately (for temperature and velocity) in upwards and downward flows, identified by the sign of $v_r - \langle v_r\rangle_\Omega$. The three different estimations of $\alpha_v$ and $\alpha_T$ are denoted by rms, up and down in Table~\ref{tab:MLT_estimation}.
Region C1 of \ag is excluded from this analysis, since that shell is not flagged as convective in the SE model at the time of mapping and therefore has no meaningful MLT prediction.

We find that in the oxygen-burning shells of both models (regions C2), the turbulent velocities exceed the MLT predictions by roughly a factor of two (see the first three rows of Table~\ref{tab:MLT_estimation}). This is consistent with previous multidimensional studies  of similar burning shells \cite{Meakin_Arnett_2007}, and corresponds to an effective mixing-length parameter of order $\alpha_{\Lambda_{\rm MLT}}^{\rm eff}\sim3$ instead of the value $1.5$ used for the SE models computed here. 
%
In these O-shells, the ratio $\alpha_{\Lambda,T}/\alpha_{\Lambda,v}$ remains order unity when rms fluctuations are used, but consistently smaller for the upward and for the downward flows (compare last three rows of Tab.~\ref{tab:MLT_estimation}) 
The values we find for $\alpha_{\Lambda,v}$ are broadly compatible with those of \cite{Meakin_Arnett_2007}, who obtained 0.84, 0.59 and 0.60 for the rms, upward, and downward flows, respectively. The larger spread of values in our models may be due to the limited  number of turnovers captured in the outer convective regions, which prevents the upward-downward asymmetry from settling into a quasi-stationary pattern. 


The thin silicon-burning region C1 of \ad behaves very differently. There the turbulent velocity is smaller in our models than the MLT prediction, and the  ratio $\left(\alpha_{\Lambda,T}/\alpha_{\Lambda,v}\right)_{\rm rms}$ deviates significantly from unity. This suggests that the shell is not well described by a standard MLT picture. The physical reason is straightforward: the silicon-burning shell is very thin,%
\footnote{An ideal-gas shell of thickness $\Delta$ located at a distance $r$ from the stellar centre becomes unstable when $\Delta/r < 1/4$ \citep{Kippenhahn_Weigert_Weiss_2013sse..book}. For C1 we find  $\Delta/r = 0.36$, close to the limiting value.}
 with a total width of only $10^8$\,cm  (Table~\ref{tab:convective_zones}), whereas the average pressure scale height of the region is $7.3\times10^7$\,cm, so the predicted mixing length exceeds the physical size of the shell. In such a configuration, the convective eddies cannot develop as assumed in the classical MLT framework--they do not fit within the shell width and are bounded above and below by two sharp mean-molecular-weight gradients. The flow is therefore more reminiscent of strongly confined turbulent convection (i.e., Rayleigh--Bénard-like convection) than of the unbounded eddy picture assumed by MLT. Supporting this interpretation,  we find a typical Rayleigh number \insr{(Eq.~\eqref{eq:Ra})} of  $10^{14}$ consistent with values typical of stellar interiors \cite{Marcus_1980}.

To characterise turbulent mixing in this thin shell, we follow \cite{Jones_2017} and reconstruct the effective 1D diffusion coefficient, $D_{1}$, that reproduces the observed mixing of chemical species in the 3D simulations. Neglecting nuclear burning, the 1D diffusion equation for a chemical species is
\begin{equation}
    \frac{\partial X_i}{\partial t} = D_1 \frac{\partial^2 X_i}{\partial x^2},
\end{equation}
To solve the previous equation, we discretise it as
\begin{equation}
\label{eqn:1D_diff_discret}
x_m \frac{X^{n+1}_k - X^{n}_k}{\Delta t}
=
D_{1,k+1}\frac{X^{n+1}_{k+1}-X^{n+1}_k}{r_{k+1}-r_k}
-
D_{1,k}\frac{X^{n+1}_k-X^{n+1}_{k-1}}{r_k-r_{k-1}} 
\end{equation}
where 
$x_m = (r_{k+1} -r_{k-1})/2$. The indices $n$ and $n+1$ correspond to two times used in the reconstruction separated by $\Delta t$, $k$ is the spatial index, and $X$ the mass fraction of a given species.\footnote{Instead of using a single isotope, we reconstruct mixing with the inverse mean molecular weight, $\mu^{-1} = \sum_i X_i(1+Z_i)/A_i$.}
We build $D_{1,k}$ from two times $t_1$ and $t_2$, averaging the composition over a temporal window of width $\pm \tau_{\rm conv, SE}$ (i.e., the temporal window corresponds to the convective turnover time defined in  Eq.~\eqref{eq:tau_conv}).
We impose $D_{1,0}=0$ just outside the convective region. 

\begin{figure}[t!]
    \centering
    \includegraphics[width=\columnwidth]{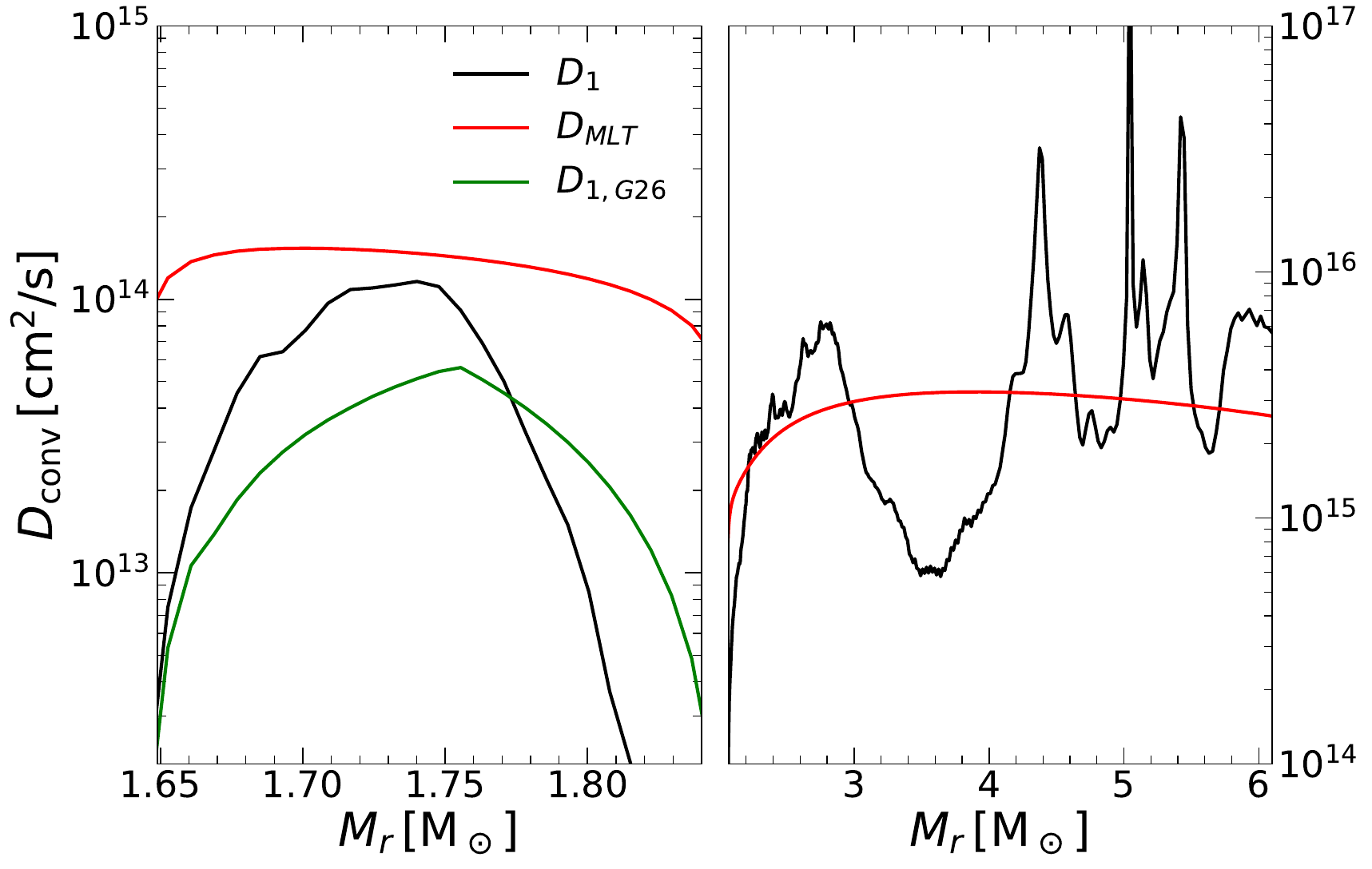}
    \caption{Diffusion coefficients for convection in region C1 (left) and C2 (right) of \ad. Estimation of the effective diffusion coefficient using Eq.~\eqref{eqn:1D_diff_discret} (black), the standard MLT prediction (red), and the corrected MLT prescription of Eq.~\eqref{eqn:D_RCMD_2} (green).}%
    \label{fig:D_conv_C1_C2}%
\end{figure}

Figure~\ref{fig:D_conv_C1_C2} compares the reconstructed diffusion coefficient with the MLT prediction for regions C1 and C2 of \ad. In the thin silicon-burning shell C1, the reconstructed profile exhibits a clear bell-like shape, with strongly reduced mixing at both shell boundaries relative to the standard MLT estimate. The peak value is slightly smaller than $D_{\rm MLT}$, consistent with the weaker turbulent velocities inferred from the 3D model.  
A similar reduction in $D_1$ near one convective boundary was identified in \cite{Jones_2017} (see their Fig.~22), who proposed a corrected prescription for the MLT diffusion coefficient
\begin{equation}
\label{eqn:D_RCMD_1}
    D_{\rm 1, J17} = v_{\rm MLT} \times \min(\alpha_{MLT} H_p, |r-r_{\rm top}|).
\end{equation}
In the present case, however, the mixing deficit occurs at both boundaries which motivates the modified prescription
\begin{equation}
\label{eqn:D_RCMD_2}
    D_{\rm 1, G26} = v_{\rm MLT} \times \min(\alpha_{MLT} H_p, min(|r-r_{\rm top}|,|r-r_{\rm bottom}|)).
\end{equation}
As shown in Fig.~\ref{fig:D_conv_C1_C2} (green line) this two sided correction reproduces the behaviour near the shell boundaries substantially better than the standard MLT diffusion coefficient. 

Applying the same reconstruction to region C2 yields a different result. The outer oxygen-burning shell does not display the same bell-shaped profile; instead, the reconstructed diffusion coefficient follows the MLT estimate reasonably well across most of the shell, with only a modest reduction near the lower boundary. On average, $D_1$ is slightly larger than $D_{\rm MLT}$, consistent with the larger turbulent velocities found in the 3D model.




\subsection{The case of a shell merger in \ad}

\begin{figure}[t!]
    \centering
    \includegraphics[width=\columnwidth]{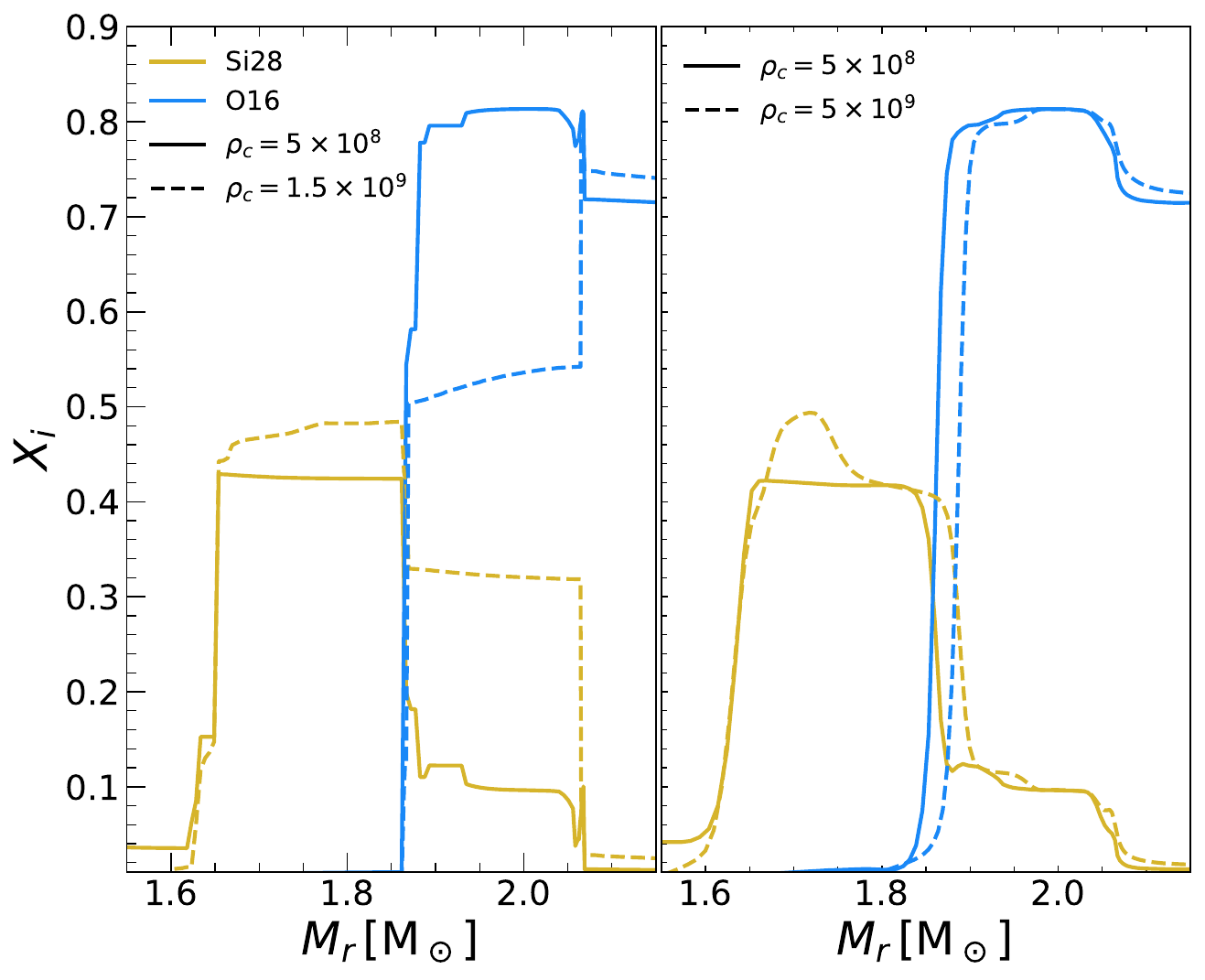}
    \caption[Mass fractions of silicon and oxygen at the start and end of the 3D simulation
 compared to before and after shell merger in the \texttt{MESA} model.]{Silicon and oxygen mass-fraction profiles from the SE calculation (left) and the shell-angle-averaged values from the 3D simulations (right). Solid lines show the initial model, and dashed lines the model immediately after the shell-merger event (for the SE calculation) and the final model before collapse (for the 3D calculation).}%
    \label{fig:Si28_O16_shell_merger}%
\end{figure}

 In the SE calculation of \ad, the oxygen shell just above C1 burns enough material during the final contraction phase for the two shells to merge just before core collapse. This merger results in a substantial outflow of silicon into the upper shell and a corresponding inflow of oxygen downwards into the silicon-burning shell, where it is rapidly processed into silicon. The left panel of Fig.~\ref{fig:Si28_O16_shell_merger} shows the silicon and oxygen mass fractions at the time of mapping and immediately after the shell-merger event in the SE model. Following the merger, the silicon mass fraction in the main shell approaches 0.5, while in the oxygen shell, extending outwards to $M_r=2.05 \,M_{\odot}$, the silicon mass fraction reaches roughly three times its initial value in that region.

In contrast, our 3D model does not show evidence for such a shell merger (see right panel of Fig.~\ref{fig:Si28_O16_shell_merger}). Oxygen continues to burn in the outer shell, but after 527\,s of evolution only a modest reduction in its mass fraction is observed. A small amount of oxygen  diffuses downward into the silicon shell, where it is converted into silicon  near the shell base, as indicated by the peak of silicon at $1.7\,M_{\odot}$. The interface between the two shells remains clearly identifiable. The 3D model may indeed be evolving toward a shell merger, but it collapses before such an event can happen. The resulting Si/O structure at collapse therefore differs substantially between the 1D and 3D models.

One likely reason for this difference is the shorter time to collapse in the multidimensional evolution. In the SE calculation, the interval between mapping and collapse is about thirty minutes, whereas in the MHD simulation it is just less than nine  minutes. The faster decrease of $Y_e$ in the 3D model (Fig.~\ref{fig:ye_evol_profile}), due to the explicit treatment of weak reactions in \texttt{RN28}, may contribute to the earlier collapse, although the final $Y_e$ difference remains modest (it is only a few percent). More generally, hydrostatic SE models contract more slowly than fully dynamical MHD models, delaying the predicted collapse time.

To assess whether the shell merger is likely to occur in the 3D model, we examine the bulk Richardson number at the Si/O interface. This quantity characterises the stiffness of the boundary between convective and radiative layers, and is often used to describe convective entrainment \citep{Meakin_Arnett_2006}. Low values of the Richardson number, $Ri_B$, imply that mixing across the boundary is easier, which can in turn further weaken the interface, potentially leading to a full shell merger.

Following \citet{Cristini_Meakin_Hirschi_Arnett_Georgy_Viallet_Walkington_2017}, we define
\begin{equation}
    \label{eqn:Bulk_Ri}
    Ri_B = \frac{\Delta B \Lambda_{Ri}}{v^2_{\rm rms}},
\end{equation}
where the buoyancy jump at the convective boundary is
\begin{equation}
    \Delta B = \int_{r_c-\Delta r}^{r_c+\Delta r} N^2 \rm dr.
\end{equation}
Here $r_c$ is the position of the interface, identified as the location of the maximum gradient in the silicon mass fraction, and $\Delta r$ is the radial extent over which the Brunt-Väisälä frequency, $N$, is integrated. For our estimation we take $\Delta r = 0.3H_p$, with $H_p$ evaluated at $r_c$, and we use the pressure scale height as the characteristic length scale $\Lambda_{Ri}$. 

The bulk Richardson number in the 3D model is larger than 200 throughout the evolution (Appendix~\ref{sec:bulk})
, implying an entrainment velocity more than a thousand times smaller than the turbulent velocity \citep{Cristini_2019}.\footnote{The estimate follows from the empirical fit $v_{\rm e}/v_{\rm rms}=A\,Ri_B^{-n}$ of \citet{Cristini_2019}, where $v_e$ is the entrainment velocity, and taking $A\approx 0.05$ and $n\approx 0.74$.} It reaches a quasi-equilibrium level by the end of the simulation, where it remains an order of magnitude higher than the corresponding SE value, of order 10, at the moment of shell merger. This strongly supports the argument that the Si/O interface remains too stiff for a  merger to occur before collapse in the multidimensional model.

\subsection{Nuclear burning}

Turbulent motions are driven by nuclear burning at the base of the convective regions, and the resulting mixing feeds back on the burning by transporting ashes upward and fresh fuel downward. In SE models, MLT accounts for this coupling provided that the mixing time scales remain shorter than the nuclear-burning time scales, allowing the two processes to be treated independently. 

\begin{figure}[t!]
    \centering
    \includegraphics[width=\columnwidth]{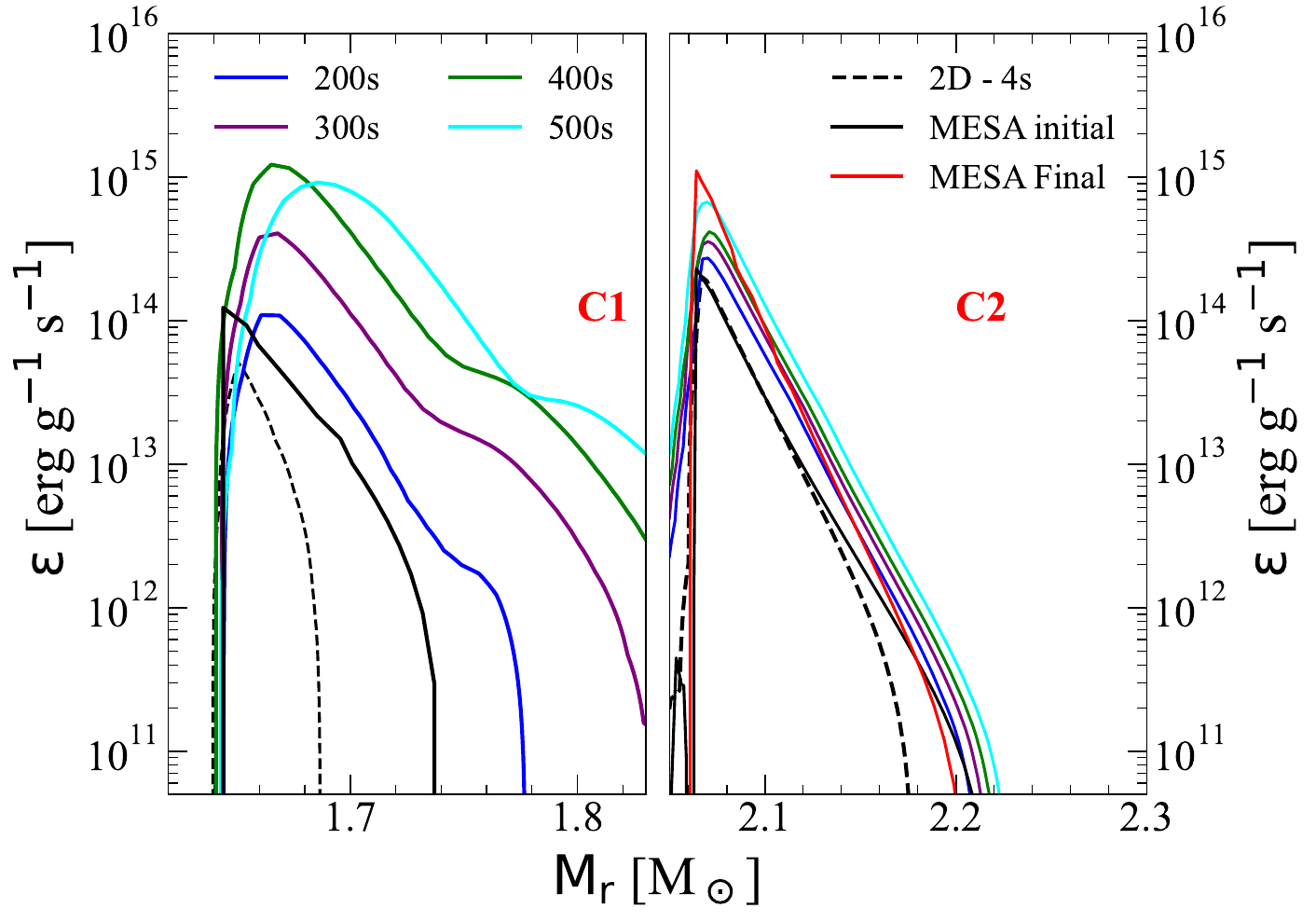}
    \caption{Total energy generation in the 3D models at different times for both nuclear-burning shells C1 (left) and C2 (right) of model \ad. The solid black line shows the energy production from the initial SE model, while the  dashed black curve corresponds to the energy generation of the 2D model after 4\,s of evolution. In the right panel, the solid red line represents the energy production from the last computed \texttt{MESA} profile.\protect\footnotemark }%
    \label{fig:etot_13M_DENA_3d}%
\end{figure}

The energy generation of our models is computed using the \texttt{RN28} network, which is essentially an $\alpha$-chain from carbon to nickel and captures the dominant nuclear energy release during the advanced burning phases.
%
Figure~\ref{fig:etot_13M_DENA_3d} compares the total net energy-generation rate --- including strong and weak reactions as well as thermal-neutrino losses --- in the multidimensional model of \ad with the corresponding 1D SE profiles. The 2D model at 4\,s after mapping (dashed black line) can be directly compared with the initial SE profile. We note a small change in the peak value in the C1 region and a faster cut off of the energy-generation profile in the multi-D model--evident, e.g., in region C1 where the sharp drop occurs at $1.69\,M_{\odot}$ instead of $1.74\,M_{\odot}$. This is a direct consequence of the explicit weak reactions included in \texttt{RN28}, which reduce the net energy release relative to the SE network. 

As the multidimensional model evolves, the burning region C1 expands in mass. This growth results from the rising temperature throughout the shell as the star contracts, combined with the downward diffusion of oxygen from the shell above (Fig.~\ref{fig:Si28_O16_shell_merger}). 
By contrast, region C2 is more stable. Here, the final SE (red line) and the final 3D (cyan line) profiles agree quite well at the peak of energy-generation rate. However, in the 3D model the burning region extends out further in mass, despite the additional loses through weak reactions. Throughout the evolution, the mass range over which the net energy-generation rate is positive increases in the 3D model, whereas in the SE model this region recedes. This behaviour reflects the stronger mixing in region C2, seen in Table~\ref{tab:MLT_estimation}, which more efficiently transports fresh fuel into the burning region and slightly enlarges it in the 3D models. Similar trends are present in region C2 of model \ag, where the final nuclear-burning region extends much further outwards.



\footnotetext{We do not show the final SE line in the left panel because, after the shell merger, the physical conditions in that region differ too strongly from those of the 3D model for the comparison to remain meaningful.}
In a convective shell close to a quasi-steady state, the convective luminosity should broadly balance the energy rate injected by nuclear burning, $\dot Q_{\rm nuc}$. This implies that the convective velocity scales as
\begin{equation}
    v^3_{\rm conv} \sim \dot{Q}_{\rm nuc} \Lambda_{\rm turb} /M_{\rm conv},
\end{equation}
where $\Lambda_{\rm turb}$ is an estimate for the coherence length of the turbulence.
Since this characteristic length is not uniquely defined, we consider two alternatives: the standard MLT estimate, $\Lambda_{\rm turb}=1.5H_p$, and the radial thickness of the convective shell, $\Lambda_{\rm turb}=\Delta_{\rm conv}$ \citep[see][for a detailed discussion]{2016_muller}. 

\begin{figure}[t!]
    \centering
    \includegraphics[width=\columnwidth]{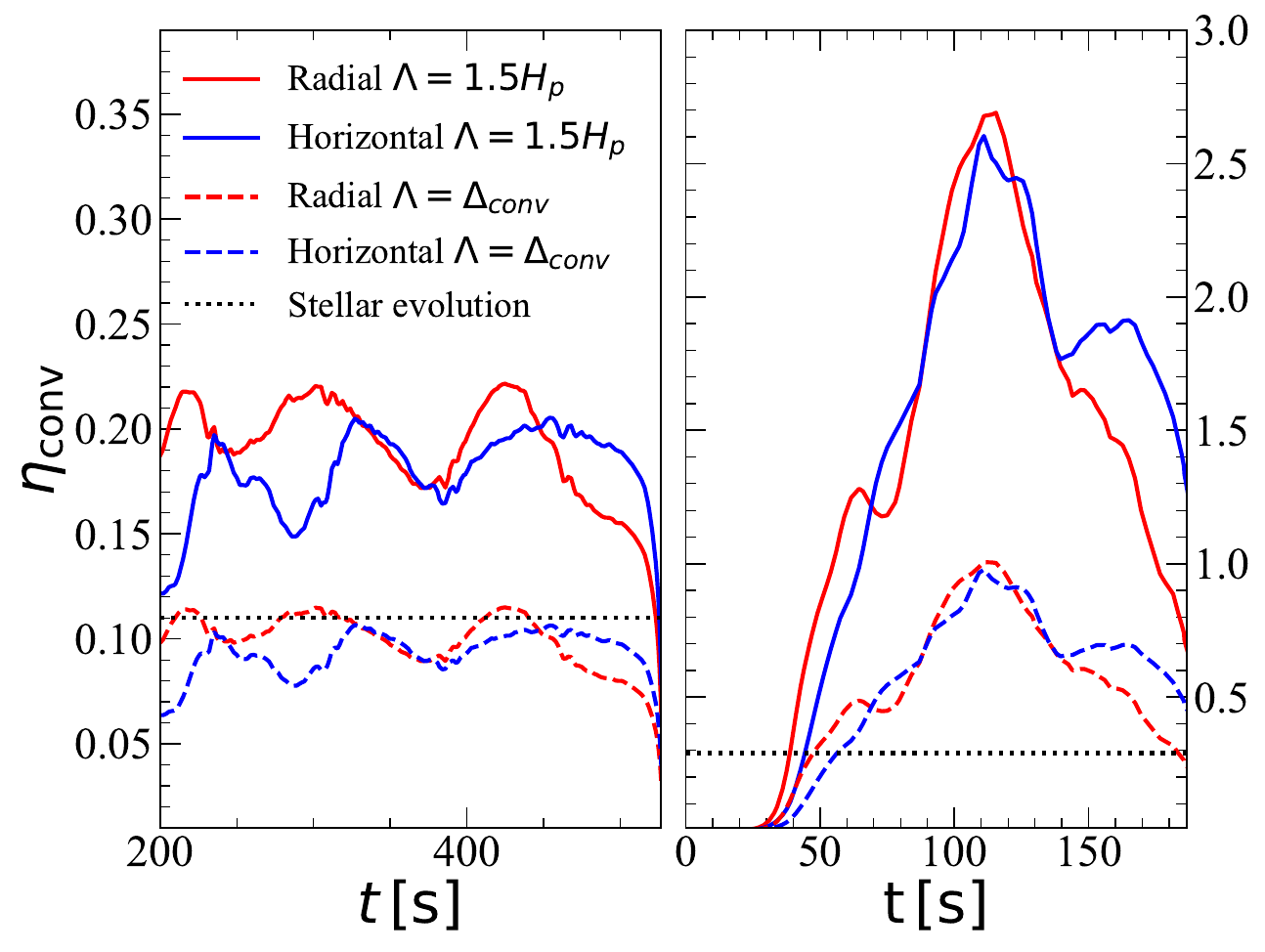}
    \caption{Conversion efficiency factor, Eq.~\eqref{eqn:eta_conv}, computed using  the characteristic length scale either $\Lambda_{\rm turb}=1.5H_p$ or the radial thickness of the burning shell. On the left we show the C2 region of model \ad and on the right the C2 region of model \ag. The SE averages  (computed with $\Lambda_{\rm turb}=1.5H_p$) are shown in black dotted lines. Values above unity indicate that the adopted $\Lambda_{\rm turb}$ is too small and/or that the shell has not yet reached a quasi-steady state.}%
    \label{fig:eta_conv_outer}%
\end{figure}

\begin{figure}[t!]
    \centering
    \includegraphics[width=0.9\columnwidth]{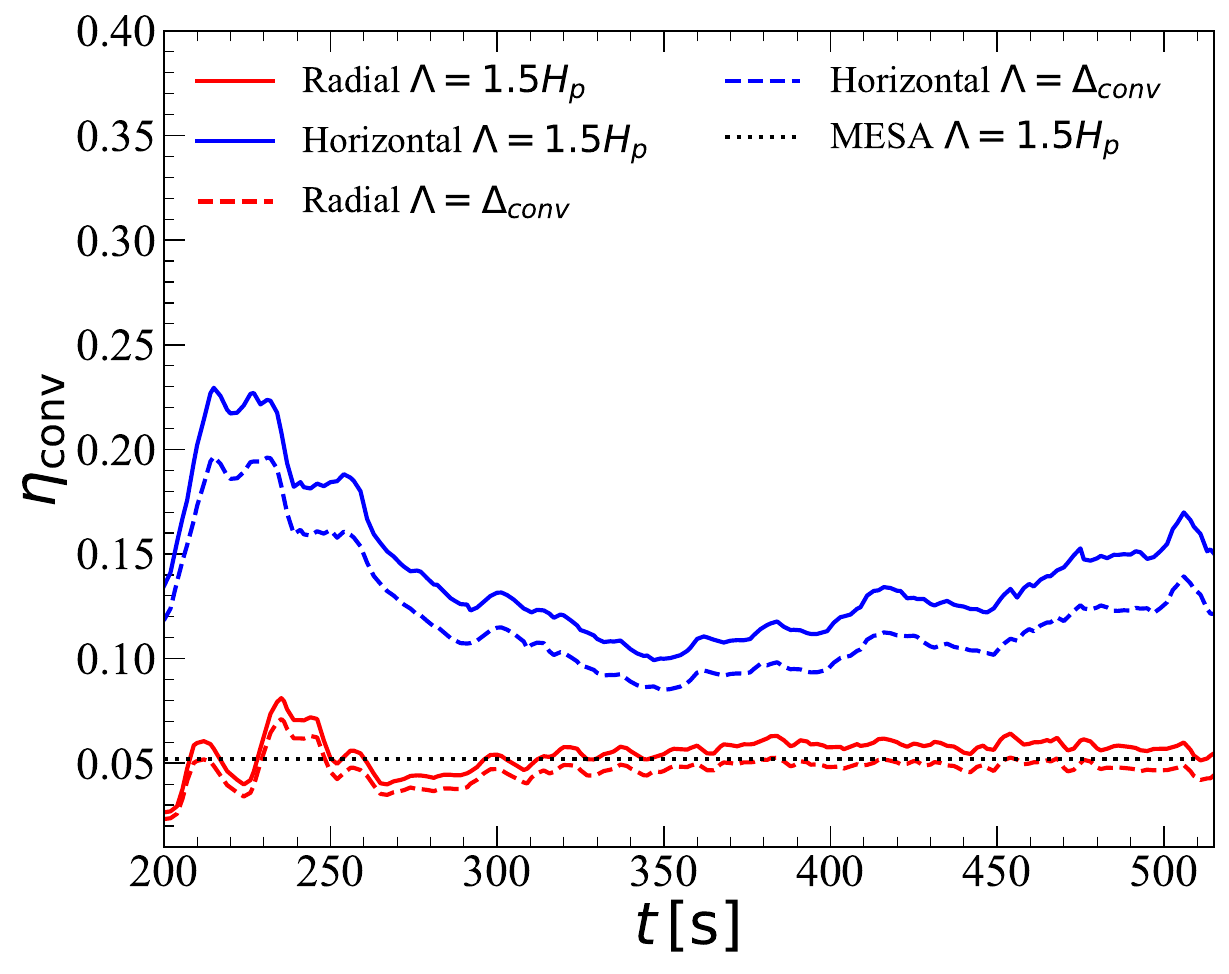}
    \caption[Conversion efficiency factor for the inner convective shell of the 3D model \ad.]{Same as Fig.~\ref{fig:eta_conv_outer} but for the zone C1 of \ad, where silicon is burning in a particularly thin shell.}%
    \label{fig:eta_conv_inner}%
\end{figure}

Following this idea, the conversion efficiency of nuclear energy into turbulent kinetic energy is defined as
\begin{equation}
\label{eqn:eta_conv}
    \eta_{\rm conv} = \frac{E_{\rm turb}/M_{\rm conv}}{\left(\dot{Q}_{\rm nuc} \Lambda_{\rm turb} /M_{\rm conv}\right)^{2/3}}.
\end{equation}
For the 3D models, we compute the turbulent kinetic energy in each convective shell separately for the radial and horizontal directions: 
\begin{align*}
    E_{\rm turb,r} &= \frac{1}{2}\int_{r_{\rm min}}^{r_{\rm max}} \!\!\!\!\!\!\mathrm{d}V
    \rho (r,\theta,\phi) \left[v_r(r,\theta,\phi) - \langle v_r ( r ) \rangle_{\Omega}\right]^2 \, , \\ 
    E_{\rm turb,h} &= \frac{1}{2}\int_{r_{\rm min}}^{r_{\rm max}} \!\!\!\!\!\!\mathrm{d}V
    \rho (r,\theta,\phi) \left[v^2_{\theta}(r,\theta,\phi) +(v_{\phi}(r,\theta,\phi) - \langle v_{\phi} ( r ) \rangle_{\Omega})^2\right]  .
\end{align*}
We then evaluate the efficiency factor for each direction (results are shown in Fig.~\ref{fig:eta_conv_outer} and Fig.~\ref{fig:eta_conv_inner}).
In the outer oxygen-burning shell C2 of \ad, the radial  and horizontal turbulent energies are close to equipartition. When $\Lambda_{\rm turb}=\Lambda_{\rm MLT}=1.5H_p$ is adopted for the multidimensional models, the values of $\eta_{\rm conv}$ are roughly twice those inferred from the SE model, even though both are evaluated with the same value of $\Lambda_{\rm turb}$. This is consistent with the stronger turbulent velocities relative to MLT reported in \mbox{Table~\ref{tab:MLT_estimation}}.
%
However, the absolute value of $\eta_{\rm conv}$ depends on the assumed coherence length. Using a larger characteristic length for the multidimensional flow lowers the inferred efficiency and brings it closer to the SE estimate, indicating that part of the difference can be interpreted as a difference in the effective turbulent coherence length rather than in the conversion efficiency alone.


The interpretation of the C2 shell of \ag is less straightforward. When $\Lambda_{\rm turb}=1.5H_p$ is used, $\eta_{\rm conv}$ becomes larger than unity and exhibits large fluctuations. This should not be taken as evidence of an anomalously efficient conversion of nuclear energy into turbulence. Rather, it suggests that $1.5H_p$ underestimates the actual coherence length of the dominant motions in this very extended shell. This interpretation is supported by the fact that adopting the shell thickness, $\Lambda_{\rm turb}=\Delta_{\rm conv}$, reduces $\eta_{\rm conv}$ to more moderate values. In addition, the large fluctuations of the efficiency correlate with the fact that this shell undergoes only a few convective turnovers during the simulation and reaches a fully turbulent state only toward the end of the run. The corresponding values of $\eta_{\rm conv}$ should therefore be regarded as provisional, since the shell is unlikely to have reached a statistically stationary state.


In the inner convective region C1 of \ad, shown in  Fig.~\ref{fig:eta_conv_inner}, the values computed using $\Lambda_{\rm turb}=\Lambda_{\rm MLT}$ or $\Lambda_{\rm turb}=\Delta_{\rm conv}$ differ only slightly, because the shell thickness is comparable to the local pressure scale height. Here the main feature is instead the anisotropy between radial and horizontal motions: the nuclear burning drives horizontal turbulence significantly more efficiently than radial turbulence, while the SE value is comparable to  the radial component.

Taken together, these results show that the multidimensional differences in turbulent transport feed back directly onto the burning structure. In the extended oxygen-burning shells, stronger turbulence broadens the burning region and increases the effective transport relative to the standard MLT description. In the thin silicon-burning shell, by contrast, confinement suppresses radial transport and lowers the radial conversion efficiency relative to the horizontal one. The shell geometry therefore affects not only the velocity field itself, but also the way in which nuclear burning sustains and shapes the turbulence.

\section{Discussion}
\label{sec:discussion}

The multidimensional calculations presented here show that the final convective shells of rapidly rotating, magnetised pre-supernova progenitors can differ from their 1D SE counterparts in both their turbulent dynamics and their burning structure. At the same time, the degree of robustness of the different results is not uniform, and it is therefore useful to distinguish between the findings that are well supported by the present simulations and those that should still be regarded as suggestive.

The most robust result concerns the thin silicon-burning shell C1 of model \ad. This stems from the fact that, for the inner most regions, the simulation time covers tens of turnover times providing our results with a high statistical significance. 
In that shell the geometry itself prevents a standard MLT-like description from being appropriate. The shell thickness is smaller than the expected mixing length, and the radial flow is strongly confined by the compositional gradients at the upper and lower boundaries. 
As a result, the turbulent flow is anisotropic, radial transport is weaker than predicted by MLT, and the effective diffusion coefficient inferred from the 3D evolution is reduced near both shell edges (see the proposed diffusion coefficient in Eq.~\eqref{eqn:D_RCMD_2}). The reconstructed diffusion profile is qualitatively similar to the boundary-limited mixing behaviour found by \citet{Jones_2017}, although in the present case the suppression occurs at both shell boundaries because of the geometric confinement of the shell.
This is also consistent with the conclusion of \citet{2016_muller} that the partition between horizontal and radial kinetic energy is not universal, but depends sensitively on the geometry of the convective region.

The behaviour of the oxygen-burning shells is also physically clear, although its quantitative interpretation is somewhat less uniform between the two models. Unlike the silicon-burning shells, the outer convective regions have longer turnover times and longer transitions times for turbulence development. Thus, the number of turnovers captured there is less significant. 
In the best-resolved cases, the multidimensional turbulent velocities exceed the standard SE estimates by approximately a factor of two. This is consistent with the oxygen-shell results of \citet{Meakin_Arnett_2007}, who likewise found convective velocities in multidimensional simulations to be larger than the corresponding one-dimensional MLT estimates. 
More broadly, it supports the view that the calibration of convective transport in advanced burning shells may differ from that adopted in standard SE models \citep{Jones_2017,Georgy_2024}. 
%

These results suggest two possible directions for improving 1D SE calculations. First, in extended oxygen-burning shells it would be worthwhile to test the impact of adopting a larger effective convective velocity, and hence a larger convective diffusion coefficient, during the final stages before collapse. The present simulations suggest that an increase by approximately a factor of two may be appropriate in at least some cases. 
%
This does not imply that standard MLT underestimates the convective velocity throughout the entire life of a massive star; rather, it suggests that a boosted velocity scale may be more appropriate for shell burning within the CO core after neon-core burning. Implementing and testing such a modification in both \GENEC and \MESA is therefore a natural next step, in order to determine whether it changes the internal structure of supernova progenitors by the time they reach collapse.

Second, in thin burning shells such as C1 in \ad, the present results indicate that the shape of the effective diffusion coefficient may be as important as its overall normalisation. In particular, the reduction of mixing near both shell boundaries appears to be a natural consequence of the shell geometry and is not captured by the standard MLT prescription. 
%
The consequences of adopting such a modified diffusion coefficient are, however, not yet straightforward to predict. In some cases, slower mixing may have little impact if the shell still has sufficient time to burn before collapse. In others, contracting shells may mix less efficiently and thus burn more slowly, leading to a different pre-supernova configuration. Such changes could affect the subsequent explosion by shifting the location of entropy jumps and by altering the final distribution of isotopes, for example through a larger surviving silicon mass fraction. 
In this sense, the present calculations support the broader programme of using multidimensional simulations to test and refine one-dimensional transport prescriptions in advanced burning stages \citep{Jones_2017,Arnett_2019,Georgy_2024}.

The shell-merger analysis of model \ad is another important result of this work. In the corresponding 1D stellar-evolution model, the silicon- and oxygen-burning shells merge during the final contraction phase before collapse. In the multidimensional MHD evolution, however, no such merger takes place before collapse. 
This does not imply that all late shell mergers predicted by SE models are artificial, but it does show that the occurrence of such mergers can be highly sensitive to the modelling framework. In the present case, the difference between the 1D and 3D evolutions leads to a significantly different Si/O structure at collapse. 
%
This has important consequences for future supernova modelling of these progenitors, since the 1D-averaged pre-SN structure of the 3D model differs substantially from that of the original 1D SE model. Our results therefore suggest that models displaying shell mergers in the final minutes prior to collapse should be used with caution as supernova progenitors, since such events may in some cases arise from limitations of the 1D treatment of the stellar interior.


More generally, the present models illustrate that multidimensional progenitors may differ from their 1D SE counterparts not only because they contain non-radial velocity perturbations, but also because their shell structure and burning geometry may evolve differently during the final minutes prior to collapse. This point is likely to be relevant for future collapse calculations, since both the amplitude and location of pre-collapse perturbations, as well as the detailed shell composition inherited by the collapsing core, can influence the subsequent dynamics \citep{2016_muller}.

Our models may also help clarify whether the location of the dominant convective shells relative to the entropy per baryon $s=4$ interface plays a role in explodability. They provide contrasting cases in which the innermost turbulent shell lies either below or above the mass coordinate $M_4\equiv M_r(s=4)$. This is particularly relevant because the quantities $M_4$ and $\mu_4$ are widely used as predictors of the fate of massive stars in one-dimensional models \citep{2016_ertl}, whereas the multidimensional structure of the flow around that interface may influence the collapse in ways that are not captured by those quantities alone.

Beyond the inclusion of rotation and magnetic fields, these progenitors also display internal structural features that have not yet been explored in supernova calculations using multidimensional initial models. They therefore provide a useful framework for assessing the limits of the 1D prescriptions adopted in SE calculations, and for identifying which of those limitations are likely to matter most for subsequent explosion modelling. In this context, the differences found in the oxygen-burning shells may ultimately have a smaller impact on the collapse and explosion than those associated with the inner convective regions, which are located closer to the core and are therefore more directly coupled to the early post-bounce dynamics.


\section{Conclusions}
\label{sec:conlusions}

We have computed the first full 3D MHD rotating supernova progenitors and followed them for several minutes prior to core collapse until reaching the pre-SN link. In this paper, we have focused on the numerical pipeline used to construct these models and on the behaviour of turbulence and nuclear burning in the shells surrounding the stellar core.

Our main results can be summarised as follows. In extended oxygen-burning shells, the turbulent velocities measured in the multidimensional simulations exceed the standard MLT predictions by about a factor of two in the best-resolved cases. This suggests that the 1D description of late shell convection may underestimate the characteristic velocity scale during the final stages of SE.

In the thin silicon-burning shell C1 of model \ad, convection is not well described by MLT. The shell is geometrically confined, the turbulent flow is strongly anisotropic, and the effective mixing is reduced near both shell boundaries. A modified 1D diffusion profile is therefore required to reproduce the behaviour seen in the 3D simulations.

The multidimensional evolution of model \ad does not exhibit the late shell merger found in the corresponding 1D SE calculation. The large bulk Richardson number at the Si/O interface indicates that the boundary remains too stiff for rapid entrainment, so that the shell structure at collapse differs substantially between the 1D and 3D models.

Finally, the differences in turbulent transport found in the multidimensional calculations feed back directly onto the burning structure of the shells. In the oxygen-burning shells, stronger mixing broadens the burning region, whereas in the thin silicon-burning shell radial transport is inhibited relative to horizontal motions. The present progenitors therefore provide not only realistic multidimensional initial conditions for future collapse studies, but also a useful benchmark for improving 1D SE modelling in the advanced burning stages.

\section{Data Availability}

The final 3D snapshots at the pre-SN link of each progenitor are stored on Zenodo \href{https://doi.org/10.5281/zenodo.19692957}{https://doi.org/10.5281/zenodo.19692957}. They will be made publicly available from 01/01/2027. For anticipated access please contact the corresponding authors.

\begin{acknowledgements}
We acknowledge support from grants PID2021-127495NB-I00 and PID2025-171322NB-C22, funded by MCIN/AEI/10.13039/501100011033 and by the European Union “NextGenerationEU". We also acknowledge support from the Astrophysics and High Energy Physics programme of the Generalitat Valenciana ASFAE/2022/026 funded by MCIN and the European Union NextGenerationEU (PRTR-C17.I1) as well as support from the Prometeo excellence programme grant CIPROM/2022/13 funded by the Generalitat Valenciana. 
\end{acknowledgements}

\bibliographystyle{aa}
\bibliography{biblio.bib}

\begin{appendix}

\section{Progenitor evolution prior to core collapse}
\label{sec:precollapse}

\begin{figure}[h!]
    \centering
    \includegraphics[width=\columnwidth]{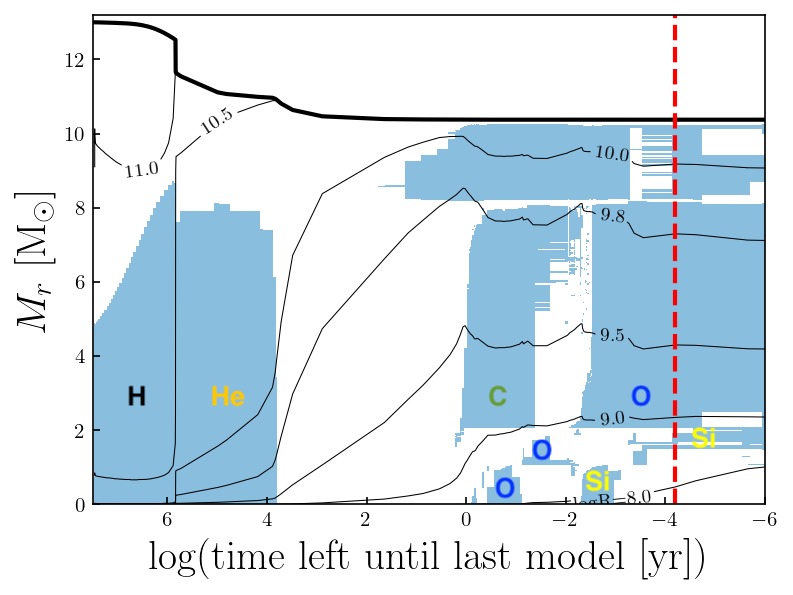}
    \includegraphics[width=\columnwidth]{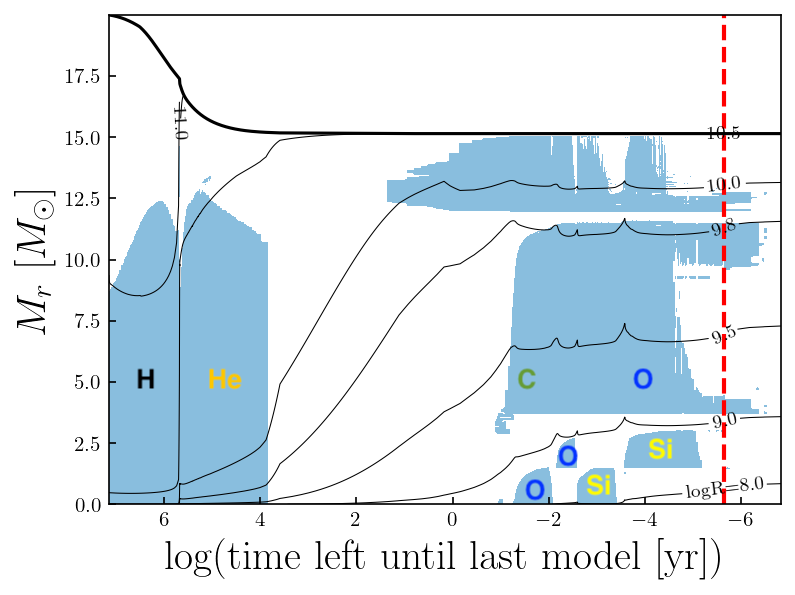}
    \caption{Kippenhahn diagram of model \ad, top, and \ag, bottom. Iso-radii
contours are shown starting at r = $10^8\,$cm until r = $10^{11}\,$cm. The thick black line shows the total mass of the star. Blue shaded areas identify regions
unstable to convection. Major burning cores and shells are labelled by
their main reacting element. Red dashed lines indicate the mapping time
from 1D SE to the subsequent 3D calculations for each model.}
    \label{fig:Kipp_13M_20M}
\end{figure}
{Figure}~\ref{fig:Kipp_13M_20M} summarises the evolutionary history of both progenitors through their Kippenhahn diagrams. Although the two models follow different evolutionary channels, they share several properties relevant for the present work: both end their lives as compact stripped-envelope stars, both develop the sequence of advanced burning shells expected prior to collapse, and both remain sufficiently compact that the multidimensional simulations can encompass almost the entire star within a reasonably resolved computational domain
(their radii are only a few $10^{10}$\,cm). 

The red dashed lines indicate the mapping times from the 1D SE calculations to the subsequent multidimensional models. In both cases, the mapping is performed during the final minutes prior to collapse, when the shell structure is already close to its pre-SN configuration. The figure also shows that, despite their broadly similar late burning histories, the detailed shell structure at mapping differs in important ways, most notably in the survival of a thin silicon-burning shell in \ad and its disappearance in \ag. We also note that after the point of mapping in \ad there is clearly interaction between the silicon and oxygen shells, which does not occur in \ag.


\section{Bulk Richardson number evolution}
\label{sec:bulk}
\begin{figure}[t!]
    \centering
    \includegraphics[width=0.85\columnwidth]{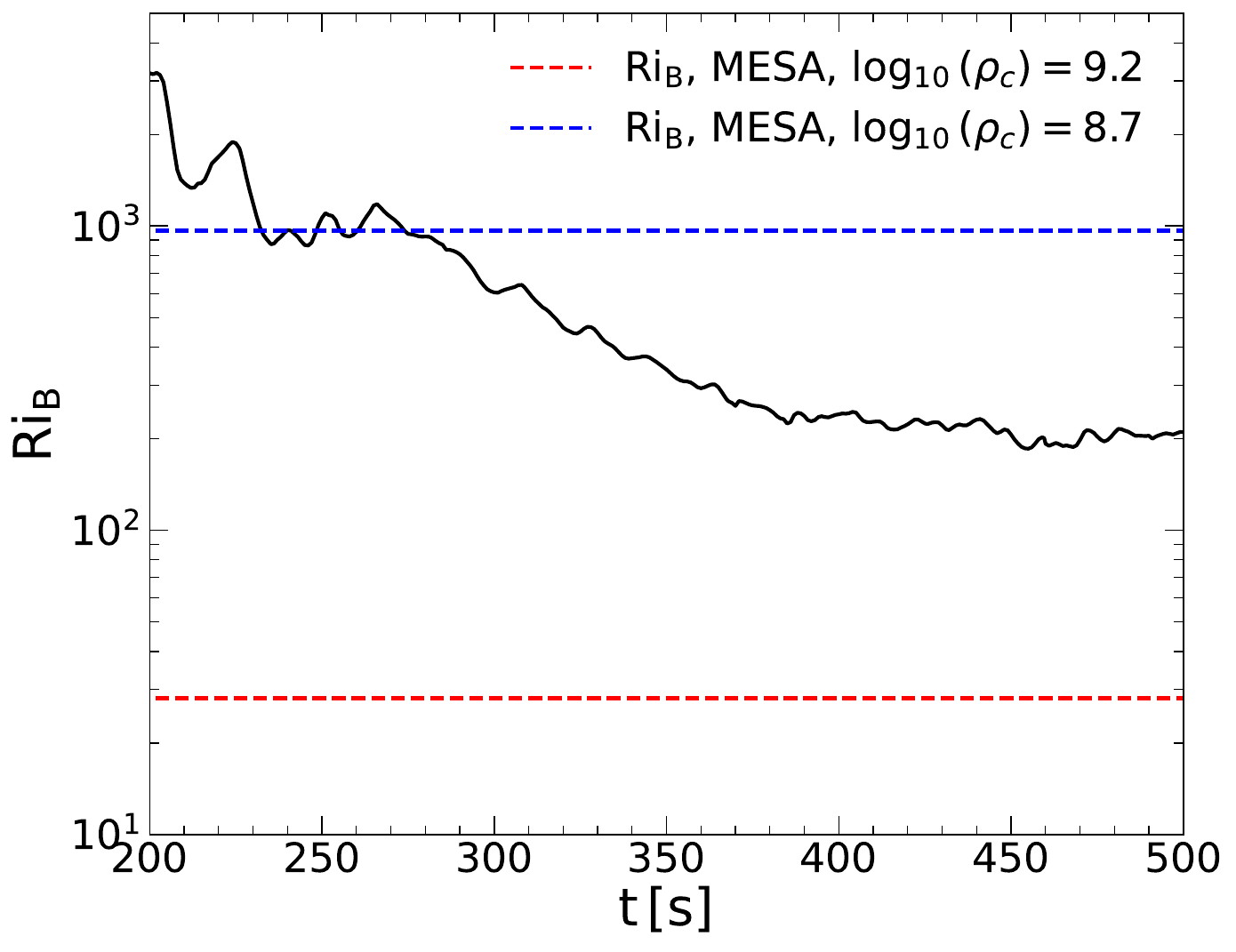}
    \caption{Evolution of the Bulk Richardson number, Eq.~\eqref{eqn:Bulk_Ri},  at the silicon–oxygen interface for the 3D model \ad. The horizontal lines correspond to the values of $Ri_B$ obtained from the SE snapshots in Fig.~\ref{fig:Si28_O16_shell_merger}.}%
    \label{fig:RiB_3D}%
\end{figure}

The bulk Richardson number (Eq.~\ref{eqn:Bulk_Ri}) is typically employed to characterise the stiffness of the boundary between convective and radiative layers. The stiffness of convective boundaries is relevant for assessing the likelihood of shell mergers. Thus, we display its evolution for the 3D model \ad in Fig.~\ref{fig:RiB_3D}, along with horizontal lines indicating the values obtained from the SE snapshots in Fig.~\ref{fig:Si28_O16_shell_merger}. Throughout the 3D evolution, $Ri_B>200$, implying an entrainment velocity more than a thousand times smaller than the turbulent velocity \citep{Cristini_2019}.\footnote{The estimate follows from the empirical fit $v_{\rm e}/v_{\rm rms}=A\,Ri_B^{-n}$ of \citet{Cristini_2019}, where $v_e$ is the entrainment velocity, and taking $A\approx 0.05$ and $n\approx 0.74$.} 
The gradual decrease in $Ri_B$ observed during the evolution reflects the increasing convective velocities as the shell heats up and burning intensifies. Nevertheless, the value reaches a quasi-equilibrium level by the end of the simulation, where it remains an order of magnitude higher than the corresponding SE value at the moment of shell merger. This strongly suggests that the Si/O interface remains too stiff for a  merger to occur before collapse in the multidimensional model. 

\section{Properties of progenitors at mapping and collapse}
\label{sec:appexdix_table}

\begin{table*}[t!]
\centering
\caption{Properties of models \ad and \ag at mapping and at collapse, as defined by the stellar-evolution calculations. All quantities are defined in Appendix~\ref{sec:appexdix_table}. The quantities $Y_{e,\rm core}$ and $M_{\rm Fe,core}$ are evaluated up to the mass coordinate where the mass fraction of $\ce{^{56}Fe}$ falls below 0.1. For each progenitor, the left column refers to the mapping model and the right column to the collapse model; for \ad, this corresponds to the snapshot at $v_r=10^8\,\mathrm{cm\,s^{-1}}$ and for \ag, to the snapshot matching the central temperature of the \ad collapse snapshot.}

\label{tab:map_and_collapse}

\begingroup
\begin{tabular}{ c c c c c c }
\hline
\hline
Models & \multicolumn{3}{c}{\ag} & \multicolumn{2}{c}{\ad} \\
\cline{2-3}\cline{5-6}

& Mapping & Collapse & & Mapping & Collapse \\

$M_{\rm tot} \ [M_{\odot}]$ & 15.1 & 15.1 & & 10.37 & 10.37 \\
$R_{\rm tot} \ [10^{10} \rm cm]$ & 2.89 & 2.89 & & 2.86 & 2.86 \\
$M_{CO} \ [M_{\odot}]$ & 11.6 & 11.6 & & 10.37 & 10.37 \\
\hline
$\rm M_{\rm Si,tot}\ [M_{\odot}]$ & 0.0376 & 0.0361 & & 0.180 & 0.209 \\
$\rm M_{\rm Fe,tot}\ [M_{\odot}]$ & 3.04 & 3.04 & & 1.65 & 1.86 \\
$\rm M_{\rm Fe,core}\ [M_{\odot}]$ & 1.50 & 1.91 & & 1.04 & 1.61 \\
\hline
$\rho_{\rm c}\ [10^8 \ \text{g\,cm}^{-3}]$ & 5.57 & 13.6 & & 4.97 & 63.1 \\
$T_{\rm c} \ [{10^9 \ \rm K}]$ & 6.01 & 7.56 & & 5.39 & 9.59 \\
$Y_{e,\rm c}$ & 0.456 & 0.451 & & 0.464 & 0.461 \\
$Y_{e,\rm core}$ & 0.468 & 0.467 & & 0.470 & 0.467 \\
$s_{\rm c} \ [\rm kb/baryon]$ & 1.06 & 1.06 & & 1.01 & 0.990 \\
\hline
$\rm \xi_{Mr=2.5}$ & 0.403 & 0.485 & & 0.209 & 0.208 \\
$\rm \xi_{s=4}$ & 0.673 & 0.840 & & 0.446 & 0.777 \\
$M_4 \ [M_{\odot}]$ & 1.50 & 1.54 & & 1.95 & 1.86 \\
$\mu_4$ & 0.188 & 0.296 & & 0.139 & 0.129 \\
\hline
\hline
\end{tabular}

\endgroup

\end{table*}


Table~\ref{tab:map_and_collapse} summarises a set of global, structural, and thermodynamic quantities for models \ag and \ad, evaluated both at the mapping point to the multidimensional simulations and at collapse, as defined by the underlying SE calculations. Since the present paper is concerned primarily with the shell structure, convective regions, and pre-supernova configuration of the models, we restrict the table to quantities that are directly relevant to those aspects. The magnetic-field and detailed rotational properties of the progenitors will be discussed in \citetalias{Griffiths2026PaperII}.

We list the total stellar mass, $M_{\rm tot}$, the total stellar radius, $R_{\rm tot}$, and the CO core mass, $M_{\rm CO}$, defined as the outermost mass coordinate for which $X_{\rm C}+X_{\rm O}>0.5$. We also provide the total silicon mass, $M_{\rm Si,tot}$, and the total iron-group mass, $M_{\rm Fe,tot}$. The inner iron-core mass, $M_{\rm Fe,core}$, is defined as the outermost mass coordinate for which the mass fraction of $\ce{^{56}Fe}$ remains above 0.1.

The central thermodynamic quantities listed are the density, $\rho_c$, temperature, $T_c$, electron fraction, $Y_{e,\rm c}$, and entropy, $s_{\rm c}$. In addition, we quote the mass-averaged electron fraction within the inner iron core,
\begin{equation}
Y_{e,\rm core} = \frac{1}{M_{\rm Fe,core}} \int_0^{M_{\rm Fe,core}} Y_e \, dm.
\end{equation}

We also list several quantities commonly used in discussions of explodability. The compactness parameter \citep{O’Connor_Ott_2011} is defined as
\begin{equation}
\label{eqn:compact}
\displaystyle \xi_{\rm Mr} = \frac{M_r/M_\odot}{R(M)/1000\,{\rm km}},
\end{equation}
and is evaluated both at $M_r=2.5\,M_\odot$ and at the mass coordinate where the entropy reaches $s=4\,k_{\rm B}\,{\rm baryon^{-1}}$. We further quote the two parameters introduced by \citet{2016_ertl},
\begin{equation}
\label{eqn:M4}
    M_4 = \left.M_r\right|_{s=4},
\end{equation}
and
\begin{equation}
\label{eqn:mu4}
    \mu_4  = \left. \frac{dm/M_\odot}{dr/1000\,{\rm km}}  \right|_{s=4}.
\end{equation}

\section{MHD equations}

\label{sec:MHD_eqs}

The multidimensional evolution is governed by the equations of compressible MHD for the density, $\rho$, velocity $\mathbf{v}$, total energy density, $e_\star$ (including the fluid and magnetic contributions), and magnetic field $\mathbf{B}$:

\begin{align}
\frac{\partial \rho}{\partial t} + \nabla \cdot (\rho \mathbf{v}) = 0, \\
\frac{\partial (\rho \mathbf{v})}{\partial t} + \nabla \cdot (\rho \mathbf{v} \otimes \mathbf{v} + \mathbf{T}) = \rho \mathbf{g}, \\
\frac{\partial e_{\star}}{\partial t} + \nabla \cdot \left[ e_{\star}\mathbf{v}+\mathbf{v}\cdot \mathbf{T}\right] = \rho (\mathbf{g}\cdot \mathbf{v} + \epsilon_{heat}),\\
\frac{\partial \mathbf{B}}{\partial t} = \nabla \times (\mathbf{v} \times \mathbf{B}),
\end{align}
where $\mathbf{g}$ is the gravitational acceleration. The stress tensor is written as
\begin{equation}
    \mathbf{T} = \left[P + \frac{\mathbf{B}^2}{2}+\rho \left(\frac{2}{3}\nu-\zeta\right)\nabla \cdot \mathbf{v}\right]\mathbf{I}-\mathbf{B} \otimes \mathbf{B} - \rho\nu\left[\nabla\otimes\mathbf{v}+(\nabla\otimes\mathbf{v})^{\rm T}\right]
    \label{eq:stress-tensor}
\end{equation}
where $P$ is the gas pressure, $\mathbf{I}$ is the identity tensor, and $\nu$ and $\zeta$ the bulk and kinematic shear viscosity respectively. 
The solenoidal constraint, $\nabla\cdot\mathbf{B}=0$, is enforced with the constrained transport method of \cite{Londrillo_2004}. 

The equations are solved in ideal, inviscid MHD, i.e., with explicitly zero resistivity ($\eta=0$) and viscosities ($\nu=\zeta=0$). Although real stellar interiors are not strictly described by ideal, inviscid MHD, the physical dissipation scales are far below the resolution of the present simulations. In practice, the unresolved magnetic and kinetic dissipation is therefore represented through the effective numerical diffusivity of the scheme \citep[see][for a full characterization of the numerical diffusivity of \Aenus]{Rembiasz_2017ApJS..230...18}. The consequences of this approximation for the magnetic-field evolution will be discussed in \citetalias{Griffiths2026PaperII}.


Our simulations do not include an explicit thermal diffusivity since all our convective regions possess a very large Péclet number. However, as an additional diagnostic of the convective regime, we estimate the effective radiative thermal diffusivity as
 \begin{equation}
\label{eqn:kappa}
\kappa_{\rm eff}=\frac{16\sigma T^3}{3\kappa_R\rho^2 c_P},
\end{equation}
where $c_P$ is the specific heat at constant pressure, $\kappa_R$ is the Rosseland mean radiative opacity, and $\sigma$ the Stefan--Boltzmann constant.

A common measure of the strength of buoyant driving relative to viscous and thermal diffusion is the Rayleigh number. For a convective layer of radial extent $\Delta_{\rm conv}$, we estimate it as
\begin{equation}
\label{eq:Ra}
    Ra = \frac{g \beta \Delta T \Delta_{\rm conv}^3}{\nu_{\rm eff}\kappa_{\rm eff}},
\end{equation}
where $\nu_{\rm eff}$ is an effective viscosity, $\Delta_{\rm conv}=r_{\rm top}-r_{\rm bottom}$ is the radial width of the convective layer, and $r_{\rm top}$ and $r_{\rm bottom}$ are the upper and lower shell boundaries. The large values obtained in the convective shells, typically $Ra\sim10^{14}$, indicate a strongly buoyancy-driven turbulent regime, consistent with the interpretation of the thin silicon-burning shell as strongly confined turbulent convection.

\section{Performance of nuclear reaction network \texttt{RN28}}

\label{sec:appendix_reac}

\begin{figure}[t!]
\centering\includegraphics[width=\columnwidth]{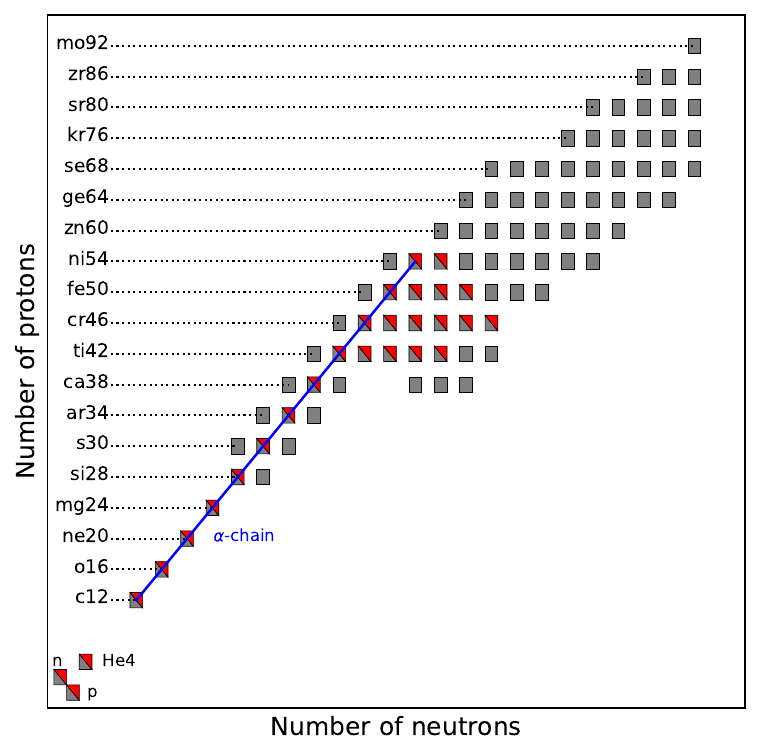} 
\caption{Species tracked by the 94-species network, \texttt{RN94} of \cite{Navo_2023} (grey) and in the reduced 28-species network \texttt{RN28} (red and grey). The blue line denotes the principal $\alpha$-chain from carbon to nickel.
}
\label{fig:networks}
\end{figure}

The multidimensional simulations presented in this work employ the reduced nuclear reaction network \texttt{RN28}. This network was designated to remain computationally affordable in 3D while still reproducing the key physics required in the pre-collapse regime, namely
the evolution of the core electron fraction  and the energetics of silicon- and oxygen-burning shells. The isotopes included in \texttt{RN28} are shown in Fig.~\ref{fig:networks}, together with those of the larger \texttt{RN94} network for comparison. 

To assess the performance of \texttt{RN28}, we performed 1D hydrodynamic simulations using both \texttt{RN28} and \texttt{RN94}, and compared the resulting evolution with that of the underlying SE models. The initial conditions correspond to the mapping models, see red lines in Fig.~\ref{fig:Kipp_13M_20M}, of \ag and \ad used in the multidimensional calculations.

\begin{figure}[t!]
\centering\includegraphics[width=\columnwidth]{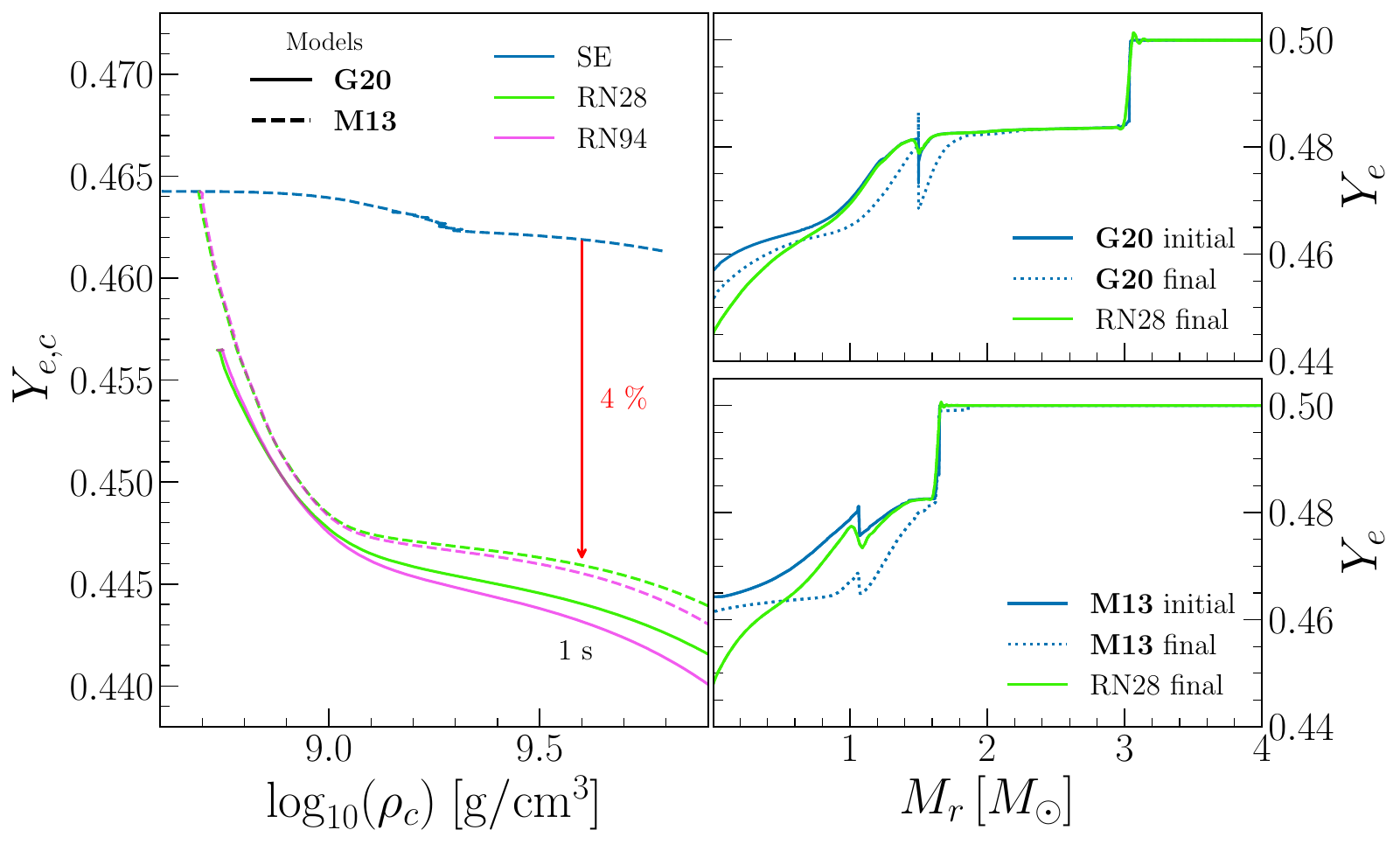} 
\caption{\textit{Left:} Evolution of the central electron fraction, $Y_{e,\rm c}$, as a function of central density for the 1D hydrodynamic models \ag and \ad, computed with \texttt{RN28} and \texttt{RN94}, together with the corresponding SE models.
\textit{Right:} Electron fraction profiles at initialisation and at collapse for \ag (top) and \ad (bottom) for the SE data and at collapse for  1D hydrodynamic model using \texttt{RN28}.}
\label{fig:ye_evol_profile}
\end{figure}

The left panel of Fig.~\ref{fig:ye_evol_profile} shows the evolution of the central electron fraction $Y_{e,\rm c}$ as a function of central density for both the SE and MHD models.

We find that the \texttt{RN28} and \texttt{RN94} networks produce nearly identical results for both models, with only slight differences emerging during the last seconds of evolution. This indicates that \texttt{RN28} captures the relevant weak-interaction physics of the larger network sufficiently well for the present application.

In contrast, both RN-based evolutions differ from the SE tracks. In particular, the hydrodynamic models exhibit an immediate change in slope with respect to the SE models in the $Y_e(\rho_c)$ plane. This difference arises because the reduced networks used in the SE calculations follow deleptonisation  through simplified electron-capture chains, whereas \texttt{RN28} and \texttt{RN94} incorporate all weak reactions among the isotopes present in each network. We therefore regard the RN-based evolution as  a more explicit representation of the weak-interaction physics during this phase. 

Despite the difference in slope, the final $Y_{e,\rm c}$ values remain relatively close. In the most extreme case, model \ad, the deviation in the final $Y_{e,\rm c}$ is only $\simeq 4\%$. The associated impact on the contraction rate is therefore likely to be modest compared with the more fundamental difference between the hydrostatic SE treatment and the hydrodynamic evolution followed here.

The right panels of Fig.~\ref{fig:ye_evol_profile} compare the electron-fraction profiles at the initial time and at collapse. We show the SE profiles together with the final hydrodynamic profile obtained with \texttt{RN28}.%
\footnote{We omit \texttt{RN94} as its results are essentially identical to those of \texttt{RN28}.}
At collapse, the central value of $Y_e$ is generally lower for model \texttt{RN28}, whereas in the outer iron core the electron fraction is slightly higher than in the SE calculation. This is likely related to the longer time interval between mapping and collapse in the SE models.%
\footnote{For instance, model \ad takes 30 minutes to evolve from the initial state to collapse in the SE calculation, whereas the corresponding hydrodynamic evolution takes only 9 minutes.} 
The longer evolution allows the approximate electron-capture chain used in the SE network to further deleptonise the core. Since this reduced EC chain can only drive $Y_e$ downward, the SE network evolves in a single direction, enhancing the discrepancy. 

Overall, these comparisons show that \texttt{RN28} reproduces the behaviour of \texttt{RN94} to very good accuracy in the regime relevant for the present work, while remaining computationally feasible for multidimensional simulations. We therefore conclude that \texttt{RN28} provides an adequate compromise between physical fidelity and numerical cost for the 3D pre-collapse calculations presented in this paper.

\end{appendix}
\end{document}